\documentclass[a4paper, 12pt]{article}
\usepackage{epsfig,amssymb,euscript,xspace, color}
\usepackage{amsmath,jheppub,mathtools}
\def\bea{\begin{eqnarray}}
\def\eea{\end{eqnarray}}
\def\be{\begin{equation}}
\def\ee{\end{equation}}



\begin{document}

\title{
On Asymptotic Symmetries \\ ~~~~~~ of 3d Extended Supergravities  }
\author{Rohan R. Poojary$^1$ {\rm and} Nemani V. Suryanarayana$^{2, \, 3}$}
\affiliation{$^1$Department of Theoretical Physics \\
~~ Tata Institute of Fundamental Research, Mumbai 400005, India\\ \\
$^2$Institute of Mathematical Sciences, \\
 Taramani, Chennai 600113, India \\ \\
$^3$Homi Bhabha National Institute,
Mumbai 400085, India}
\emailAdd{ronp@theory.tifr.res.in, nemani@imsc.res.in}

\abstract{We study asymptotic symmetry algebras for classes of three dimensional supergravities with and without cosmological constant. In the first part we generalise some of the non-dirichlet boundary conditions of $AdS_3$ gravity to extended supergravity theories, and compute their asymptotic symmetries. In particular, we show that the boundary conditions proposed to holographically describe the chiral induced gravity and Liouville gravity do admit extension to the supergravity contexts with appropriate superalgebras as their asymptotic symmetry algebras. In the second part we consider generalisation of the 3d BMS computation to extended supergravities without cosmological constant, and show that their asymptotic symmetry algebras provide examples of nonlinear extended superalgebras containing the $BMS_3$ algebra.}

\maketitle

\section{Introduction}
In the standard statement of the AdS/CFT correspondence as a duality between a gravitational theory in $AdS_{d+1}$ spacetime and a $CFT_d$, the definition of the bulk theory requires specifying boundary conditions for all its fields near the boundary of the $AdS_{d+1}$ space. The most extensively used boundary conditions for the bulk metric are the ones first proposed by Brown and Henneaux (BH) \cite{Brown:1986nw} -- which consist of holding the boundary metric fixed and specifying a certain fall-off conditions of the metric components away from the boundary. Then the asymptotic symmetry algebra of BH is obtained by studying the asymptotic Killing vector fields and the algebra of the corresponding charges. In particular, for the $AdS_3$ gravity the BH boundary conditions give rise to the 2-dimensional conformal algebra - two commuting copies of the Virasoro algebra - as the asymptotic symmetry algebra with the central charge $c=3\ell/2G$. Their results have since been generalised to include $AdS_3$ supergravites \cite{deBoer:1998kjm, Ito:1998vd, Henneaux:1999ib} where people have uncovered 2-dimensional superconformal agebras as the corresponding asymptotic symmetry algebras. The results thereof enabled successful embedding of BH boundary conditions into string theory context with applications in AdS/CFT. The 2d superconformal algebras with extended supersymmetry are generically nonlinear. The computations of \cite{deBoer:1998kjm, Ito:1998vd, Henneaux:1999ib} thus have also provided realizations of such nonlinear conformal superalgebras as asymptotic symmetry algebras of $AdS_3$ supergravities.

However, BH boundary conditions are not the only admissible ones for the $AdS_3$ gravity. In recent times it has been shown (see for instance, \cite{Banados:2002ey, Compere:2008us, Troessaert:2013fma, Compere:2013bya, Avery:2013dja, Apolo:2014tua, rrp2015thesis, Grumiller:2016pqb, Perez:2016vqo, Krishnan:2016dgy} etc.) that non-dirichlet boundary conditions for the metric also lead to consistent sets of boundary conditions of $AdS_3$ gravity which also admit appropriately defined asymptotic symmetry algebras that are infinite dimensional.

The non-dirichlet boundary conditions of $AdS_3$ gravity would involve bulk configurations which have some (if not all) components of the boundary metric fluctuating. In fact one can construct a diffeomorphism invariant theory by simply coupling the CFT to a dynamical background metric minimally. Such diffeomorphic theory will also exhibit Weyl symmetry classically. This Weyl invariance in general may not survive quantisation. An example being the string worldsheet theory where demanding vanishing Weyl anomaly restricts the matter sector of the theory - this then allows one to gauge fix the 2d worldsheet metric completely. When the Weyl anomaly survives the metric cannot be completely gauged away and there will be one component left dynamical in the metric. The problem of quantising such a gravitational theory was fist addressed by Ployakov \cite{Polyakov:1987zb, Polyakov:1981rd}.
    
The effective theory one obtains by integrating over the original 2d CFT matter content is non-local in terms of the metric and is termed as an induced gravity theory. Diffeomorphism invariance can be used to bring the metric into either of the two standard forms in terms of which the induced gravity action becomes local, namely: (i) the light-cone gauge $ds^2=-dx^+dx^-+F(x^+,x^-)(dx^+)^2$ and (ii) the conformal gauge $ds^2=-e^{\phi(x^+, \, x^-)}dx^+dx^-$.

Polyakov studied the 2d induced gravity theory in the ligh-cone gauge, namely, the Chiral induced gravity (CIG), and uncovered a hidden $sl(2,\mathbb{R})$ symmetry. Motivated to describe such chiral induced gravity holographically we  proposed in \cite{Avery:2013dja} (generalised and studied further in \cite{Apolo:2014tua}) a set of boundary conditions for $AdS_3$ gravity which admitted an asymptotic symmetry algebra that included an $sl(2, {\mathbb R})$ current algebra. Soon after Polaykov's work it was generalised to include supersymmetry in the 2d theory (see \cite{Grisaru:1987pf, HariDass:1988dp}). Also to be able to embed the CIG boundary conditions of \cite{Avery:2013dja} into some string theory context one needs to generalise them to supergravity contexts first. One of the aims of this current paper is to provide supersymmetric generalisation of \cite{Avery:2013dja}.

One can work in the conformal gauge for the 2d induced gravity and this leads to the Liouville theory. To model such theory holographically one has to let the conformal factor of the boundary metric to fluctuate. In \cite{Troessaert:2013fma} Troessaert provided the first example of such boundary conditions. In this paper we embed the boundary conditions of \cite{Troessaert:2013fma} also into supergravity contexts and thus show that they are also admissible by supersymmetry. In this case we find that generically the asymptotic symmetry algebra generalises to nonlinear superalgebra extending the cases of \cite{Henneaux:1999ib} and \cite{Troessaert:2013fma}.

In \cite{Troessaert:2013fma} imposed an additional condition (as it was done for the light-cone gauge in \cite{Compere:2013bya}) that the boundary metric has vanishing scalar curvature -- making the conformal factor to satisfy the free field equation -- and not the Liouville equation. Just as \cite{Avery:2013dja} considers relaxation of the Ricci flatness of the boundary metric of \cite{Compere:2013bya}, one can relax the boundary conditions of \cite{Troessaert:2013fma} so that the boundary metric has non-zero (constant) scalar curvature. It is then easy \cite{rrp2015thesis} to see that the bulk equations of motion match that of Liouville equation. In the appendix \ref{liouville} to this paper we study the asymptotic symmetry algebra of this case too both in the second order and first order formulations of $AdS_3$ gravity.

In the second part of the paper we change tracks and consider supersymmetric generalisation of BMS boundary conditions \cite{Bondi:1962px, Sachs:1962zza} for flat 3d gravity without cosmological constant (${\mathbb R}^{1,2}$ gravity). Some of the important examples of this case have been considered in \cite{Banados:2004nr, Barnich:2014cwa, Barnich:2012aw, Banerjee:2016nio, Banerjee:2017gzj}. Building on these results, we show that for every case in \cite{Henneaux:1999ib} of the extended $AdS_3$ supergravity with the corresponding superalgebra as its asymptotic symmetry algebra there is a corresponding asymptotic supersymmetry algebra in the flat case. It is well known that the asymptotic supersymmetry algebra for extended $AdS_3$ supergravities are typically nonlinear. We show that the corresponding asymptotic supersymmetry algebras of ${\mathbb R}^{1,2}$ supergravity are also nonlinear, thus providing first examples of nonlinear super extensions of $BMS_3$ algebras.

This paper is organized as follows: In section \ref{sugra} we extend the chiral boundary conditions of \cite{Avery:2013dja, Apolo:2014tua} to the supersymmetric setting - first to the simplest minimal supergravities and then to extended supergravities. In section \ref{troessaert} we embed the results  of \cite{Troessaert:2013fma} to extended supergravities. In section \ref{Flat_sugra} we take a suitable flat limit of the results in \cite{Henneaux:1999ib} and obtain $BMS_3$ embedded in extended asymptotic superalgebra. In Appendix \ref{liouville} we study supersymmetric extensions of conformal boundary conditions with non-vanishing boundary curvature. This will further be analysed in the Chern-Simons formulation of 3d gravity. We conclude with discussion of our results in section 6.  

\section{Holographic Chiral Induced Supergravities}
\label{sugra}
Motivated to describe the 2-dimensional Chiral Induced Gravity of Polyakov holographically, in the second order formulation of $AdS_3$ gravity the following boundary conditions \cite{Avery:2013dja, Apolo:2014tua} were proposed for the metric
\begin{equation}
\label{apsbdyconds}
\begin{aligned}
g_{rr} &= \frac{l^2}{r^2} + {\cal O}(r^{-4}), ~~ g_{r+} = {\cal O}(r^{-1}), ~~ g_{r-} = {\cal O}(r^{-3}), \\
g_{+-} &= - \frac{r^2}{2} + {\cal O}(r^0), ~~ g_{--} =  {\cal O}(r^0), \\
g_{++} &= r^2 F(x^+, x^-) + {\cal O} (r^0) ,
\end{aligned}
\end{equation}
where $x^+, x^-$ are the boundary coordinates and $r$ is the radial coordinate with the asymptotic boundary at $r^{-1} = 0$. After imposing the conditions coming from variational principle that allowed the function $F(x^+, x^-)$ to fluctuate, it was shown in \cite{Avery:2013dja} that the asymptotic symmetry algebra consists of one copy of Virasoro and one copy of $sl(2, {\mathbb R})$ current algebra with level $k$ given by $l/4G$. Those boundary conditions were translated to the first order Chern-Simons (CS) formulation of $AdS_3$ gravity in \cite{Poojary:2014ifa}.

In this section we generalise the CIG boundary conditions of \cite{Avery:2013dja, Apolo:2014tua} to supersymmetric contexts. 

\subsection{ ${\mathcal N}=(1,1)$ Supergravity in $AdS_3$}
We first begin with a simple set-up where we generalise the chiral boundary conditions of \cite{Avery:2013dja}. We will work in the Chern-Simons (CS) formulation as the calculations are simpler in this formulation of supergravity in $AdS_3$. The graded Lie algebra of interest here is $osp(1|2)$ which contains in it the bosonic $sl(2,{\mathbb R})$. The commutation relations  are as follows:
\bea
\left[ \sigma^0,R^{\pm} \right] &=& \pm\tfrac{1}{2}R^{\pm },\cr
\left[ \sigma^0,\sigma^{\pm} \right]&=&\pm \sigma^{\pm},\hspace{0.9cm}\left[ \sigma^{\pm},R^{\pm} \right]=0,\cr
\left[ \sigma^+,\sigma^- \right]&=&2\sigma^0,\hspace{1cm} \left[ \sigma^{\pm},R^{\mp} \right]=R^{\pm},\cr
\left\{ R^{\pm},R^{\pm} \right\}&=&\pm\sigma^{\pm},\hspace{0.9cm} \left\{ R^{\pm},R^{\mp} \right\}=-\sigma^0, \cr
\sigma^0&=&\tfrac{1}{2}\sigma^3,\hspace{1cm}\left[ \sigma^0,R^{\pm} \right]=\pm\tfrac{1}{2}R^{\pm }.
\eea
The gauge-invariant, bi-linear, non-degenerate metric on the algebra is:
\be
Tr(\sigma^a \sigma^b)=h^{ab}=\frac{1}{4}
\begin{pmatrix}
2&0&0\\
0&0&4\\
0&4&0
\end{pmatrix},\hspace{1cm}STr\left( R^{-}R^{+} \right)=-STr\left( R^{+}R^{-} \right)=1.
\ee
The ${\mathcal N}=(1,1)$ supergravity action can be written as a difference to two Chern-Simons actions \cite{Achucarro:1987vz, Witten:1988hc}, 
\bea
S_{sugra-AdS_3}&=&S_{CS}[\Gamma]-S_{CS}[\tilde{\Gamma}],\cr
S_{CS}[\Gamma]&=&\frac{k}{4\pi}\int STr( \Gamma\wedge d\Gamma +\tfrac{2}{3}\Gamma\wedge \Gamma\wedge \Gamma),\cr
{\rm where}\,\,\,\Gamma&=&\left[ A_{a\mu}\sigma^a  +\psi_{+\mu}R^+ +\psi_{-\mu}R^- \right]dx^\mu
\eea
where the gauge algebra for the two CS terms is $osp(1|2)$.\footnote{The product of two fermions differs by a factor of $i$ from the standard Grasmann product ($(ab)^*=b^*a^*$); this  amounts to using $STr\left( R^{-}R^{+} \right)=-STr\left( R^{+}R^{-} \right)=-i$ and  $\left\{ R^{\pm},R^{\pm} \right\}=\mp i\sigma^{\pm},\,\,\left\{ R^{\pm},R^{\mp} \right\}=i\sigma^0.$} 
The equation of motion imposes flatness condition on the two gauge fields valued in the adjoint of $osp(1|2)$.
In order to obtain the required generalization, we first notice the form of the gauge fields corresponding to the chiral boundary conditions written down in \cite{Avery:2013dja, Poojary:2014ifa}. 
We take them to be the following gauge fields 
\bea
\label{APSCSansatz}
a&=&\left[ L_1-\kappa L_{-1} \right]dx^+,~~~ \tilde{a} =\left[ -L_{-1}+\tilde{\kappa}L_1 \right]dx^-+ f^{(a)}L_a dx^+,\cr
A&=&b^{-1}(d+a)b, ~~~ \tilde{A} = b(d+\tilde{a})b^{-1},
\eea
with $b = e^{\ln(r/\ell) L_0}$. It is easy to check that these also yield the same chiral boundary conditions as in \cite{Avery:2013dja} on the metric but the metric will no longer be in Fefferman-Graham gauge. Therefore these would also correspond to the chiral boundary conditions studied therein. 

Here we note that the fluctuating field at the boundary comes from the $f^{(-1)}$ component of $\tilde{a}_+$ in (\ref{APSCSansatz}). The components of $\tilde{a}_-$ play the role of sources, {\it i.e.,} functions that need to be specified like a chemical potential. We further notice that all the $\tilde{a}_+$ components are $a\,\, priori$ turned on and are determined in terms of $f^{(-1)}$. On the other hand, the $\tilde{a}_-$ component with leading $r$ dependence is fixed to be $-1$, while only the sub-leading component, $L_1$ of $\tilde{a}_-$ is allowed to have coordinate dependence.

Therefore taking a cue from the above observation, we propose the following fall-off conditions for the supersymmetric case:
\bea
\Gamma &=& bdb^{-1}+bab^{-1},\cr
\tilde{\Gamma}&=& b^{-1}db + b^{-1}\tilde{a}b,\cr
{\rm where}\,\,\,b&=&e^{\sigma^0\ln(r/\ell)},\cr
a&=&\left[  \sigma^- + L \, \sigma^+ +\psi_+R^+     \right]dx^+,\cr
\tilde{a}&=&\left[\sigma^+ + \bar{L} \, \sigma^-+\bar{\psi}_-R^- \right]dx^-+\left[ \tilde{A}_{a+}\sigma^a  + \tilde{\psi}_+ R^++ \tilde{\psi}_- R^- \right]dx^+ \cr
&:=& \tilde a_- dx^- + \tilde a_+ dx^+
\eea
where we have relabelled the generators as $L_{\pm}=\pm \sigma^{\mp}$ and $L_0=\sigma^0$. Here the $dx^-$ component of the gauge field 1-form $\tilde{a}$  is that of a super-gauge field corresponding to Dirichlet boundary condition as given in \cite{Banados:1998pi,Bautier:1999ds}. All functions above are ${\it a\, priori}$ functions of both the boundary coordinates. The equations of motion imply flatness:
\bea
\partial_+\Gamma_--\partial_-\Gamma_++[\Gamma_+,\Gamma_-] = 0, ~~ \partial_+\tilde{\Gamma}_--\partial_-\tilde{\Gamma}_++[\tilde{\Gamma}_+,\tilde{\Gamma}_-] = 0.
\eea
 For the left gauge field $a$ this implies that the functions are independent of the $x^-$ coordinate. $i.e.$ $\partial_-a=0$. 
\be
\partial_-L=\partial_-\psi_+=0.
\ee
While for the right gauge field $\tilde a$ components, equation of motion allows one to express the $dx^+$ components of $\tilde{a}$ in terms of $dx^-$ components:
\bea
\tilde{A}_{0+}&=&\partial_-\tilde{A}_{++}, ~~~ \tilde{A}_{-+} = \tilde{A}_{++}\bar{L} - \tfrac{1}{2}\partial^2_-\tilde{A}_{++}+\tfrac{i}{2}\tilde{\psi}_+\bar{\psi}_-,\cr
\tilde{\psi}_-&=&\tilde{A}_{++}\bar{\psi}_--\partial_-\tilde{\psi}_+ .
\eea 
The remaining relations imposed by equations of motion are differential equations relating the $dx^+$ components and  the $dx^-$ components of $\tilde{a}$. These are interpreted as Ward identities for the boundary theory:
\bea
\partial_+\bar{L}+\tfrac{1}{2}\partial^3_-\tilde{A}_{++}-2\bar{L}\partial_-\tilde{A}_{++}-
\tilde{A}_{++}\partial_-\bar{L}
+i\bar{\psi}_-\left( \tilde{A}_{++}\bar{\psi}_- +  \partial_-\tilde{\psi}_+ \right)\! &+&\!i \partial_-(\bar{\psi}_-\tilde{\psi}_+)=0,\cr
\partial_+\bar{\psi}_- -\partial_-[\tilde{A}_{++}\bar{\psi}_- - \partial_-\tilde{\psi}_+ ] - \tfrac{1}{2}\partial_-\tilde{A}_{++}\bar{\psi}_- -\bar{L}\tilde{\psi}_+\! &=& \!0.
\label{n1ward}
\eea
Here, conventionally (according to Brown-Henneaux analysis) the $\tilde{A}_{++},\tilde{\psi}_+$, are sources $i.e.$ chemical potentials coupling to conserved currents labelled by  $\bar{L},\bar{\psi}_-$ respectively. But our boundary conditions would require that the currents $\bar{L},\bar{\psi}_-$ play the role of sources. 
We will later choose these sources such that global $AdS_3$, corresponding to $\bar{L}=-\tfrac{1}{4}$ and $\bar{\psi}=0$, is part of the space of bulk solutions. 

The boundary terms required to make the set of solutions with fixed $\bar{L}$ and  $\bar{\psi}$ variationally well defined are:
\bea
S_{bndy}&=&\frac{k}{8\pi}\int_{\mathcal \partial M}d^2x\,\, STr(-\sigma^0[\tilde{a}_+,\tilde{a}_-]-2\bar{L}_{(0)}\sigma^-\tilde{a}_+ -\tfrac{1}{2}\bar{\psi}_{(0)-} \, R^-\tilde{a}_+ ).
\eea
The variation of the total action therefore reads:
\bea
\delta S_{total}&=&\frac{k}{8\pi}\int_{\mathcal M}\!\!d^2x\,\, \left[2(\bar{L}-\bar{L}_{(0)})\delta\tilde{A}_{++}+\tfrac{i}{2}(\bar{\psi}_-
-\bar{\psi}_{(0)-})\delta\tilde{\psi}_+ \right]
\eea
Here, choosing the fluctuations $\delta\tilde{A}_{++}$ and $\delta\tilde{\psi}_+$ to vanish would impose Dirichlet type boundary condition, whereas allowing for their fluctuations demands setting $\bar{L}=\bar{L}_0$ and $\bar{\psi}_-=\bar{\psi}_0$ to satisfy the variational principal. We choose the later case as this implies fields fluctuating on the boundary of $AdS_3$ as we seek. 

One may try and generalise the boundary conditions of Comp\`ere $et\,\,al$ \cite{Compere:2013bya} by choosing only $x^+$ dependence for $\tilde{A}_{++}$ for arbitrary values of the sources $\bar{L}_0$ and $\bar{\psi}_{(0)-}$; but on solving the above Ward identities, one sees that $\tilde{\psi}_+$ cannot just be a function of $x^+$. Therefore allowing  $\tilde{\psi}_+$ to depend on $x^-$ leads to generalisation of the boundary conditions of \cite{Compere:2013bya} to the supergravity case. As we will see this case can be thought of as a special case of more general analysis of the next subsection \ref{extendedsugra}. 
\subsubsection{Charges and Symmetry Algebra}
We first solve the equation of motion ($\ref{n1ward}$) for a particular value of $\bar{L}=-\tfrac{1}{4}$ and $\bar{\psi}_-=0$. This choice of $\bar{L}$ allows for global $AdS_3$ to be one of the allowed solutions. This implies that the boundary fields $\tilde{A}_{++}$ and $\tilde{\psi}_+$ take the following form: 
\bea
\tilde{A}_{++}&=&f(x^+)+g(x^+)e^{ix^-}+\bar{g}(x^+)e^{-ix^-},\cr
\tilde{\psi}_+&=&\chi(x^+)e^{ix^-/2}+\bar{\chi}(x^+)e^{-ix^-/2}.
\label{N11sol}
\eea
The residual gauge transformations that leave $\tilde{a}$ form-invariant are:
\bea
&&\tilde{\Lambda}=\xi_a\sigma^a  + \epsilon_{\pm}R^{\pm}, ~~~
\delta\tilde{a} =d \tilde{\Lambda} +[\tilde{a},\tilde{\Lambda}],\cr&&\cr
\implies&&\xi_0=\partial_-\xi_+, ~~ \xi_- = -\tfrac{1}{4}(1+2\partial_-^2)\xi_+, ~~~
\epsilon_- = -\partial_-\epsilon_+,\cr
&&\partial_-(1+\partial^2_-)\xi_+=0=(\partial^2_-+\tfrac{1}{4})\epsilon_+.
\eea
One can solve for the residual gauge transformation parameters:
\bea
\xi_+&=&\lambda_f(x^+)+\lambda_g(x^+)e^{ix^-}+\bar{\lambda}_{\bar{g}}(x^+)e^{-ix^-},\cr
\epsilon_+&=&\varepsilon(x^+)e^{ix^-/2}+\bar{\varepsilon} (x^+)e^{-ix^-/2}.
\eea 
The left gauge field components are independent of $x^-$ and the  corresponding residual gauge transformations are parametrised by:
\bea
\Lambda&=&\zeta_a\sigma^a+\varepsilon_\pm R^\pm,\cr
\delta a&=&d\Lambda+\left[ a,\Lambda \right],\cr
\zeta_0&=&-\partial_+\zeta_-,\cr
\zeta_+&=&-\tfrac{1}{2}\partial^2_+\zeta_-+\zeta_-L-i\psi_+\varepsilon_-,\cr
\varepsilon_+&=&-\partial_+\varepsilon_-+\zeta_-\psi_+.,\cr
0&=&\partial_-\zeta_-=\partial_-\varepsilon_-,
\eea
where all the parameters are determined in terms of $\zeta_-(x^+)$ and $\varepsilon_-(x^+)$. 
We observe that the arbitrary functions specifying the space of gauge fields $a$ and $\tilde{a}$ and the space of residual gauge transformations are specified by functions of $x^+$ alone.
Therefore the $x^+$ dependence of the functions will be suppressed here onwards. The variation of the above parameters under the residual gauge transformations for the right sector are:
\bea
\delta f&=&\lambda_f' + 2i(g\bar{\lambda}_{\bar{g}}-\bar{g}\lambda_g)+i(\chi\bar{\varepsilon}+\bar{\chi}\varepsilon),\cr
\delta g&=&\lambda_g' +i(g\lambda_f-\lambda_g f)+i\chi\varepsilon,\cr
\delta\bar{g}&=&\bar{\lambda}_{\bar{g}}' -i(\bar{g}\lambda_f-\bar{\lambda}_{\bar{g}}f)+i\bar{\chi}\bar{\varepsilon},\cr
\delta\chi&=&\varepsilon'  + i[g\bar{\varepsilon} - \tfrac{1}{2}f\varepsilon - \lambda_g\bar{\chi}+\tfrac{1}{2}\lambda_f\chi],\cr
\delta\bar{\chi}&=&\bar{\varepsilon}' -i[\bar{g}\varepsilon - \tfrac{1}{2}f\bar{\varepsilon}- \bar{\lambda}_{\bar{g}}\chi +\tfrac{1}{2}\lambda_f\bar{\chi}]
\eea 
Similarly those for the left sector are:
\bea
\delta L&=&-\frac{1}{2}\zeta_-'''+\left[(\zeta_-L)'+\zeta_-'L  \right]-i\left[ \tfrac{1}{2}(\psi_+\varepsilon_-)'+\psi_+\varepsilon'_- \right]\cr
\delta \psi_+&=&-\varepsilon''_-+\left[ (\zeta_-\psi_+)'+\tfrac{1}{2}\zeta'_-\psi_+ \right]+L\varepsilon_-.
\eea
The charges corresponding to these transformation (which can be computed using the formalism of \cite{Barnich:2001jy, Barnich:2007bf}) are given by \cite{Henneaux:1999ib}:
\bea
\mathrlap{\slash}\delta Q [\Lambda,\tilde{\Lambda}]&=& \tfrac{k}{2\pi}\int d\phi\,\, \left\{ Str[\Lambda,\delta a_\phi] - Str[\tilde{\Lambda},\delta\tilde{a}_\phi]\right\}.
\eea
The above charge can be integrated and is finite. The charges for the two gauge fields decouple:
\bea
Q[\tilde{\Lambda}]&=&-\tfrac{k}{2\pi}\int d\phi ~ [-\tfrac{1}{2}f\lambda_f + g\bar{\lambda}_{\bar{g}} + \bar{g}\lambda_g  + \chi\bar{\varepsilon}-\bar{\chi}\varepsilon  ],\cr
Q\left[ \Lambda \right]&=&\tfrac{k}{2\pi}\int \! d\phi \left(  \zeta_- L+i\varepsilon_-\delta\psi_+\right).
\eea
This charge is the generator of canonical transformations on the space of solutions parametrized by set of functions $F$ $via$ the Poisson bracket.
\bea
\delta_\Lambda F &=& \{Q[\Lambda],F \}
\eea
Therefore the Poisson bracket algebra for the right sector has to be:
\bea
&&\{f({x^+}'),f(x^+)\}=-2\alpha\delta'({x^+}'-x^+),\hspace{1.5cm}\{\chi ({x^+}'),f(x^+) \}=-i\alpha \delta({x^+}'-x^+)\chi,\cr
&&\{g({x^+}'),f(x^+)\}=-2i\alpha g(x^+)\delta({x^+}'-x^+),\hspace{0.6cm}\{\bar{\chi}({x^+}'),f(x^+) \}=i\alpha \delta({x^+}'-x^+)\bar{\chi},\cr
&&\{\bar{g}({x^+}'),f(x^+)\}=2i\alpha \bar{g}({x^+})\delta({x^+}'-x^+),\hspace{0.9cm}\{\bar{\chi} ({x^+}'),g(x^+) \}=i\alpha \delta({x^+}'-x^+)\chi,\cr
&&\{\bar{g}({x^+}'),g(x^+)\}=i\alpha f(x^+)\delta({x^+}'-x^+),\hspace{1.1cm}\{\chi ({x^+}'),\bar{g}(x^+) \}=-i\alpha \delta({x^+}'-x^+)\bar{\chi}.\cr
&&\hspace{2.8cm}+\alpha \delta'({x^+}'-x^+)
\eea
While the fermionic Poisson brackets are:
\bea
&&\{\bar{\chi}({x^+}'),\chi(x^+)\}=\tfrac{i\alpha}{2}f(x^+)\delta({x^+}'-x^+)+\alpha\delta'({x^+}'-x^+),\cr
&&\{\chi({x^+}'),\chi(x^+)\}=i\alpha g(x^+)\delta({x^+}'-x^+),\cr
&&\{\bar{\chi}({x^+}'),\bar{\chi}(x^+)\}=i\alpha\bar{g}(x^+)\delta({x^+}'-x^+).
\eea
where $\alpha =\tfrac{2\pi}{k}$. Rescaling the above currents to:
\bea
&&f\rightarrow\tfrac{k}{4\pi}f,\hspace{0.4cm}g\rightarrow\tfrac{k}{2\pi}g,\hspace{0.4cm}\bar{g}\rightarrow\tfrac{k}{2\pi}\bar{g},\hspace{0.4cm}\chi_{\alpha}\rightarrow\tfrac{k}{2\pi}\chi_{\alpha},\hspace{0.4cm}\bar{\chi}_{\alpha}\rightarrow\tfrac{k}{2\pi}\bar{\chi}_{\alpha},
\eea
and expanding them in the Fourier modes in $x^+$ yields the following commutators:
\bea
&&[f_m,f_n]=m\tfrac{k}{2}\delta_{m+n},\hspace{2.4cm}[\chi_m,f_n]=\tfrac{1}{2}\chi_{m+n},\cr
&&[g_m,f_n]=g_{m+n},\hspace{2.9cm}[\bar{\chi}_m,f_n]=-\tfrac{1}{2}\bar{\chi}_{m+n},\cr
&&[\bar{g}_m,f_n]=-\bar{g}_{m+n},\hspace{2.6cm}[\bar{\chi}_m,g_n]=-\chi_{m+n},\cr
&&[\bar{g}_m,g_n]=-2f_{m+n}-mk\delta_{m+n,0},\hspace{0.5cm}[\chi_m,\bar{g}_n]=\bar{\chi}_{m+n},
\eea
and anti-commutators:
\bea
&&\{\chi_m,\chi_n\}=-g_{m+n+2a},\hspace{2.2cm}\{\bar{\chi}_m,\bar{\chi}_n\}=-\bar{g}_{m+n+2a},\cr
&&\{\bar{\chi}_m,\chi_n\}=-f_{m+n+2a}-k(m+a)\delta_{m+n+2a,0}.
\eea
This yields the familiar affine $sl(2,{\mathbb R})$ current algebra at level $k=c/6$ with two additional fermionic current parametrized by $\chi,\bar{\chi}$.  Here $a=0$ is the NS sector and $a=1/2$ is the Ramond sector\footnote{This can be seen from the form of solution (\ref{N11sol}) where there is a factor of $e^{\pm ix/2}$ multiplying each fermionic mode}. Of these only the NS sector admits a maximal finite dimensional sub-superalgebra. These fermionic currents form a semi-direct sum with the $sl(2,{\mathbb R})$ current algebra elements with their anti-commutators yielding the latter. 

Similarly the left sector yields the familiar Brown-Henneaux result \cite{Bautier:1999ds}. We first redefine the currents by suitable scaling:
\be
L\rightarrow\tfrac{k}{2\pi}L\,\,,\,\,\psi_+\rightarrow\tfrac{k}{2\pi}\psi_+.
\ee
After these redefinitions one gets the following Poisson algebra:
\bea
\left\{ L(x'^+),L(x^+) \right\}&=&\frac{k}{4\pi}\delta'''(x'^+-x^+)-\left( L(x'^+-)+L(x^+) \right)\delta'(x'^+-x^+),\cr
i\left\{ \psi_+(x'^+),\psi_+(x^+) \right\}&=&-\tfrac{k}{\pi}\delta''(x'^+-x^+)+2L(x'^+)\delta(x'^+-x^+),\cr
\left\{ L(x'^+),\psi_+(x^+) \right\}&=&-\left[ \psi_+(x'^+)+\tfrac{1}{2}\psi_+(x^+) \right]\delta'(x'^ +-x^+).
\eea
The Fourier modes for the above algebra satisfy the following Dirac brackets:
\bea
\left[ L_m,L_n \right]&=&(m-n)L_{m+n}+\tfrac{k}{2}m^3\delta_{m+n,0},\cr
\left\{ \psi_{+m},\psi_{+n} \right\}&=&2L_{m+n+2a}+2k(m+a)^2\delta_{m+n+2a,0},\cr
\left[ L_m,\psi_{+n} \right]&=&(\tfrac{m}{2}-n)\psi_{+(m+n)},
\eea
Here unlike the right sector $a=0$ implies Ramond and $a=1/2$ implies the NS boundary condition for the fermions. This is the Virasoro algebra with an affine super-current parametrised by $\psi_+$. 

\subsection{Generalization to Extended $AdS_3$ Supergravity}
\label{extendedsugra}
Next, we will generalise the analysis of the last subsection to extended supergravity setting with negative cosmological constant. Since the number for gauge field components would now increase to include the ones corresponding to the internal bosonic directions, it would be interesting to see whether the chiral boundary conditions proposed admit a unique non-trivial generalisation. That is, we would seek boundary fall-off conditions for the gauge field components such that all of them admit a fluctuating mode at the $AdS_3$ boundary with the solutions elucidated in the pure gravity case \cite{Avery:2013dja} being a subset. The ${\mathcal N}=(4,4)$ case which would be of interest for realising these boundary conditions in a string theoretic setting would therefore be a special case of this analysis. 

One first begins by classifying the superalgebras possible in $AdS_3$ \cite{Nahm:1977tg, Henneaux:1999ib}. Let G denote the graded Lie algebra, such that $G=G_0\oplus G_1$, where $G_0$ denotes the even part whereas $G_1$ denotes the odd part. The even part, $G_0$ must contain a direct sum of $sl(2,{\mathbb R})$ and an internal symmetry algebra denoted by $\tilde{G}$.  The fermions must transform in the 2-dimensional spinor representation of $sl(2,{\mathbb R})$. The dimension of the internal algebra $\tilde{G}$ is denoted by $D$ while the fermions transform under a representation $\rho$ ($dim\, \rho=d$) of $\tilde{G}$ which is real but not necessarily unitary. This is possible in a set of seven cases, whose list can be found for example in \cite{Henneaux:1999ib}.\footnote{We again follow the conventions of \cite{Henneaux:1999ib} which are summarised in the Appendix \ref{Appendixextendedsugra}.}

The gauge field is then written as a super gauge field valued in the adjoint of any of the allowed (graded Lie) superalgebra:
\bea
\Gamma&=&\left[ A_{a\mu}\sigma^a +B_{a\mu}T^a+ \psi_{+\alpha\mu}R^{+\alpha} +\psi_{-\alpha\mu}R^{-\alpha} \right]dx^\mu \, .
\eea 
The gauge field as written above, separates as a sum of $sl(2,{\mathbb R})$, $\tilde{G}$ and fermionic 1-forms. The parameters $A_a$ and $B_a$ commute\footnote{It is understood in the above context that since $A_a$ parametrizes the gauge one-form along the $sl(2,{\mathbb R})$, its index $a$ runs from $\{0,+,-\}$. While the index $a$ for  the parameter $B_a$ runs from $\{1,\cdots, D\}$ along the internal $\tilde{G}$ basis.}, while  $\psi_{\pm\alpha}$ are anti-commuting Grasmann parameters. The gauge field $\Gamma$ is a $G$ valued 1-form.

The supergravity action is then given as the difference of two Chern-Simons action at level $k$ written for two such gauge fields $\Gamma$ and $\tilde{\Gamma}$.
\be
S[\Gamma,\tilde{\Gamma}]=S_{CS}[\Gamma]-S_{CS}[\tilde{\Gamma}].
\ee 
We will concern ourselves with the case where both $\Gamma$ and $\tilde{\Gamma}$ are valued in the same $G$. The cases where this is not so leads to chiral action and can be regarded as one of the ways to generate chiral asymptotic symmetries.
\subsubsection{Boundary conditions}
Now, we would like to impose boundary conditions on the gauge fields- just as in the higher-spin case \cite{Poojary:2014ifa}, which generalize the chiral boundary conditions on pure $AdS_3$ of \cite{Avery:2013dja}. Here we would allow one of the super-gauge fields- $\Gamma$, to obey the boundary conditions of \cite{Henneaux:1999ib} $i.e.$ consistent with Brown-Henneaux, while proposing new boundary conditions on $\tilde{\Gamma}$. The BH type boundary conditions on $\Gamma$ are already analysed in \cite{Henneaux:1999ib} and so we skip the details here. The chiral boundary conditions on $\tilde \Gamma$ generalised to extended supergravity will be analysed in detail below.

The fall-off conditions in terms of the gauge fields are:
\bea
\Gamma &=& bdb^{-1}+bab^{-1}, ~~~ \tilde{\Gamma}= b^{-1}db + b^{-1}\tilde{a}b,\cr
a&=&\left[  \sigma^- + L\sigma^+ +\psi_{+\alpha}R^{+\alpha} + B_{a+}T^a    \right]dx^+,\cr
\tilde{a}&=&\left[\sigma^+ + \bar{L}\sigma^-+\bar{\psi}_{-\alpha}R^{-\alpha}+\bar{B}_{a-} T^a \right]dx^-+\left[ \tilde{A}_{a+}\sigma^a + \tilde{B}_{a+}T^a + \tilde{\psi}_{\pm\alpha}R^{\pm\alpha} \right]dx^+.\nonumber\\
\eea
where $b=e^{\sigma^0\ln(r/\ell)}$, refer to Appendix \ref{Appendixextendedsugraboundary} for more details. The flatness of the super-connection $\tilde \Gamma$ requires that the functions therein satisfy the following set of differential equations:
\bea
\partial_+\bar{L}+\tfrac{1}{2}\partial^3_-\tilde{A}_{++}-2\bar{L}\partial_-\tilde{A}_{++}
-\tilde{A}_{++}\partial_-\bar{L}
&&\cr+i\eta^{\alpha\beta}\bar{\psi}_{-\beta}\left( \tilde{A}_{++}\bar{\psi}_{-\alpha} + (\lambda^a)^\beta_{\,\,\alpha}\bar{B}_{a-}\tilde{\psi}_{+\beta}  + \partial_-\tilde{\psi}_{+\alpha} \right) +i \eta^{\alpha\beta}\partial_-(\bar{\psi}_{-\beta}\tilde{\psi}_{+\alpha})&=&0,\cr
\partial_+\bar{B}_{a-}-\partial_-\tilde{B}_{a+}+f^{bc}_{\,\,\,\,\,a}\tilde{B}_{b+}\bar{B}_{c-}+i\tfrac{d-1}{2C_\rho}(\lambda^a)^{\alpha\beta}\tilde{\psi}_{+\alpha}\bar{\psi}_{-\beta}&=&0,\cr
\partial_+\bar{\psi}_{-\alpha} -\partial_-[\tilde{A}_{++}\bar{\psi}_{-\alpha} - \partial_-\tilde{\psi}_{+\alpha} + (\lambda^a)^\beta_{\,\,\alpha}\bar{B}_{a-}\tilde{\psi}_{+\beta}] - \tfrac{1}{2}\partial_-\tilde{A}_{++}\bar{\psi}_{-\alpha} &&\cr+ (\lambda^a)^\beta_{\,\,\alpha}\bar{B}_{a-}[\tilde{A}_{++}\bar{\psi}_{-\beta} - \partial_-\tilde{\psi}_{+\beta} + (\lambda^a)^\gamma_{\,\,\beta}\bar{B}_{b}\tilde{\psi}_{+\gamma}] -(\lambda^a)^\beta_\alpha\tilde{B}_{a+}\bar{\psi}_{-\beta}
-\bar{L}\tilde{\psi}_{+\alpha}&=&0.
\eea
These are the Ward identities expected to be satisfied by the induced supergravity theory on the boundary. 

Generalising the minimal case of the previous subsection we choose boundary conditions such that global $AdS_3$ with $\bar{L}=-\tfrac{1}{4}$ and $\bar{B}=0=\bar{\psi}$ is a part of the space of bulk solutions. The boundary term to be added so as to make the right sector with fixed $\bar{L}$, $\bar{B}$ and $\bar{\psi}$  variationally well-defined is given by:
\bea
S_{bndy}&=&\frac{k}{8\pi}\int_{\mathcal \partial M}d^2x\,\, \Big[STr(-\sigma^0[\tilde{a}_+,\tilde{a}_-]-2\bar{L}_{(0)}\sigma^-\tilde{a}_+ + (\tfrac{d-1}{2C_\rho})^2T^aT^bSTr(\tilde{a}_+T_a)STr(\tilde{a}_-T_b)\cr
&&\hspace{1.5cm} -2(\tfrac{d-1}{2C_\rho})\bar{B}_{(0)a}T^aT^bSTr(\tilde{a}_+T^b)-\tfrac{1}{2}\bar{\psi}_{(0)-\alpha}R^{-\alpha}\tilde{a}_+ ) \Big].
\eea
This results in the following desired variation of the total action:
\bea
\delta S_{total}&=&\frac{k}{8\pi}\int_{\mathcal M}\!\!d^2x\,\, \Big[2(\bar{L}-\bar{L}_{(0)})\delta\tilde{A}_{++}+2(\tfrac{2C_\rho}{d-1})(\bar{B}_{a-}-\bar{B}_{(0)a})\delta\tilde{B}_{a+}\cr&&\hspace{1.7cm}+\tfrac{i}{2}(\bar{\psi}_{-\alpha}
-\bar{\psi}_{(0)-\alpha})\delta\tilde{\psi}_{+\alpha}\eta^{\alpha\beta} \Big]
\eea
Here, one has an option of choosing $\delta\tilde{()}$ functions to vanish at the $AdS$ asymptote, implying a Brown-Henneaux type boundary condition where $\tilde{A}_{++},\tilde{B}_{a+},\tilde{\psi}_{+\alpha}$ act as chemical potentials. Or, alternatively, treat $\bar{L}_{(0)},\bar{B}_{(0)a},\bar{\psi}_{(0)-\alpha}$ as chemical potentials allowing $\tilde{A}_{++},\tilde{B}_{a+},\tilde{\psi}_{+\alpha}$ to fluctuate - thus describing a theory of induced gravity on the boundary. In our present case, we choose the later by fixing $\bar{L}_{(0)}=-1/4,\bar{B}_{(0)a}=0=\bar{\psi}_{(0)-\alpha}$. Thus the variational principle is satisfied for configurations with $\bar{L}=-\tfrac{1}{4}$ and $\bar{B}_{a-}=0=\bar{\psi}_{-\alpha}$ which describes global $AdS_3$.
\subsubsection{Charges and Symmetry Algebra}
Just as in the previous subsection, one needs to find the space of gauge transformations that maintains the above form of the gauge fields, thus inducing transformations on the functions $\tilde{A}_{a+},\tilde{B}_{a+},\tilde{\psi}_{+\alpha},L,B_a,\psi_{+\alpha}$ which parametrize the space of solutions. Once this is achieved, one can define asymptotic conserved charge associated with the change induced by such residual gauge transformations on the space of solutions.
\vskip .5cm
\noindent\underline{\bf The Left Sector}:
\vskip .5cm
The analysis of the left sector is exactly as in \cite{Henneaux:1999ib} which we skip giving the details of here. One basically gets the super-Virasoro with quadratic nonlinearities as the asymptotic algebra for the modes of parameters labelling the left sector gauge filed $\Gamma$.

\bea
\left[ L_m,L_n \right]&=&(m-n)L_{m+n}+\tfrac{k}{2}m^3\delta_{m+n,0},\cr
\left[ B^a_m,B^b_n \right]&=&-f^{abc}B^c_{m+n}+\tfrac{2kC_\rho}{d-1}m\delta^{ab}\delta_{m+n,0},\cr
\left[ L_m,B^a_n \right]&=&-nB^a_{m+n},\cr
\left\{ (\psi_{+\alpha})_m,(\psi_{+\beta})_n \right\}&=&2\eta_{\alpha\beta}L_{m+n+2a}-2i\tfrac{d-1}{2C_\rho}(m-n)(\lambda^a)_{\alpha\beta}(B_a)_{m+n+2a}\cr
&+&2k\eta_{\alpha\beta}(m+a)^2\delta_{m+n+2a,0}\cr
&-&k(\tfrac{d-1}{2kC_\rho})^2\left[ \left\{ \lambda^a,\lambda^b \right\}_{\alpha\beta}+\tfrac{2C_\rho}{d-1}\eta_{\alpha\beta}\delta^{ab} \right](B_aB_b)_{m+n+2a},\cr
\left[ L_m,(\psi_{+\alpha})_n \right]&=&(\tfrac{m}{2}-n)(\psi_{+\alpha})_{m+n},\cr
\left[ B^a_m,(\psi_{+\alpha})_n \right]&=&i(\lambda^a)^\beta_{\,\,\alpha}(\psi_{+\beta})_{m+n},
\eea
with $a=0$ being Ramond and $a=1/2$ being NS boundary conditions on the fermions.
This is the nonlinear super-conformal algebra. The central extension is $k=c/6$, and is the same for all the seven cases listed in the table of \cite{Henneaux:1999ib} mentioned previously. This algebra, although a supersymmetric extension of the Virasoro algebra, is not a graded Lie algebra in the sense that the right hand sides of the fermionic (Rarita-Schwinger) anti-commutators contain quadratic nonlinearities in currents for the internal symmetry directions.
 
\vskip .5cm
\noindent\underline{\bf The Right Sector}:
\vskip .5cm

The analysis of the right sector is similar to the one covered in the ${\mathcal N}=(1,1)$ case whose details can be found in the Appendix \ref{Appendixextendedsugraright}. Here we choose 
$\bar{L}=-\tfrac{1}{4},\bar{B}=0=\bar{\psi}$ as the values for the chemical potential as it would allow for global $AdS_3$ as one of the solutions.
The asymptotic symmetry algebra in Fourier modes of the parameters labelling the right gauge field $\tilde{\Gamma}$ is
 
\bea
&&[f_m,f_n]=m\tfrac{k}{2}\delta_{m+n,0},\hspace{4.1cm}[(\chi_\alpha)_m,f_n]=\tfrac{1}{2}(\chi_\alpha)_{(m+n)},\cr
&&[g_m,f_n]=g_{m+n},\hspace{4.9cm}[(\bar{\chi}_\alpha)_m,f_n]=-\tfrac{1}{2}(\bar{\chi}_\alpha)_{(m+n)},\cr
&&[\bar{g}_m,f_n]=-\bar{g}_{m+n},\hspace{4.6cm}[(\bar{\chi}_\alpha)_m,g_n]=-(\chi_\alpha)_{m+n},\cr
&&[\bar{g}_m,g_n]=-2f_{m+n}-mk\delta_{m+n,0},\hspace{2.2cm}[(\chi_\alpha)_m,\bar{g}_n]=(\bar{\chi}_\alpha)_{m+n},\cr&&\cr
&&\{(\chi_\alpha)_m,(\chi_\beta)_n\}=-\eta_{\alpha\beta}g_{m+n+2a},\hspace{2.2cm}\{(\bar{\chi_\alpha})_m,(\bar{\chi_\beta})_n\}=-\eta_{\alpha\beta}\bar{g}_{m+n+2a},\cr
&&[(\tilde{B}_{a+})_m,(\chi_\beta)_n]=-(\tfrac{d-1}{2C_\rho})(\lambda_a)^\alpha_{\,\,\,\beta}(\chi_\alpha)_{(m+n)},\hspace{0.4cm}[(\tilde{B}_{a+})_m,(\bar{\chi}_\beta)_n]=-(\tfrac{d-1}{2C_\rho})(\lambda_a)^\alpha_{\,\,\,\beta}(\bar{\chi}_\alpha)_{(m+n)},\cr&&\cr
&&[(\tilde{B}_{a+})_m,(\tilde{B}_{b+})_n]=-i(\tfrac{d-1}{2C_\rho})f_{ab}^{\,\,\,\,c}(\tilde{B}_{c+})_{(m+n)}-(\tfrac{d-1}{2C_\rho})km\delta_{ab}\delta_{m+n,0},\cr
&&\{(\bar{\chi}_\alpha)_m,(\chi_\beta)_n\}=-\eta_{\alpha\beta}f_{(m+n+2a)}+i(\lambda^a)_{\alpha\beta}(\tilde{B}_{c+})_{(m+n+2a)}-k(m+a)\eta_{\alpha\beta}\delta_{m+n+2a,0},
\eea
with the reversed identification of $a=0$ corresponding to NS and $a=1/2$ corresponding to Ramond boundary conditions on the fermions. This is the affine Ka\v{c}-Moody super-algebra. Here, it is evident that the central extension to the $sl(2,{\mathbb R})$-current sub-algebra spanned by ($f,g,\bar{g}$) is $k=c/6$. The quadratic nonlinearities that occur in the super-Virasoro are not present here. Also, as in the ${\cal N}=(1,1)$ case the right \& left sector with NS  boundary conditions on the fermions give a maximal global subalgebra.

Thus demanding that one considers all types of fields $(\tilde{A}, \tilde B,\tilde{\psi})$ have fluctuating components on the boundary of asymptotic $AdS_3$, we have constructed a unique generalisation of the boundary condition studied in \cite{Avery:2013dja} to extended supergravity in asymptotically $AdS_3$ spaces. In doing so we uncovered the expected super-Virasoro algebra with quadratic nonlinearities for the left sector and a Ka\v{c}-Moody super-current algebra at level $k=c/6$ for the right sector. Here, we have demanded as before that the global $AdS_3$ remains in the space of allowed solutions. 

There exists a consistent truncation of this superalgebra with $g$ and $\bar{g}$ set to zero. This would correspond to a supergravity generalisation of Comp\`ere $et\,\,al$ \cite{Compere:2013bya}. In that case the residual gauge transformations could be appropriately chosen so that they do not turn on $g$ and $\bar{g}$\footnote{Also, a different choice of $\bar{L},\bar{B}\,\&\,\bar{\psi}_{-\alpha}$ can also be made}. It then corresponds to a super-extension of the $u(1)$ Ka\v{c}-Moody algebra of \cite{Compere:2013bya}.
\section{Holographic induced super-Liouville theory}
\label{troessaert}
To describe the 2-dimensional Induced Gravity in the conformal gauge holographically one would start with boundary conditions of $AdS_3$ gravity that allow the conformal factor of the boundary metric to fluctuate. Such boundary conditions proposed by Troessaert \cite{Troessaert:2013fma} look like:
\begin{equation}
\begin{aligned}
g_{rr} &= \frac{l^2}{r^2} + {\cal O}(r^{-4}), ~~ g_{r+} = {\cal O}(r^{-1}), ~~ g_{r-} = {\cal O}(r^{-1}), \\
g_{+-} &= - \frac{r^2}{2}F(x^+, x^-) + {\cal O}(r^0), ~~ g_{--} =  {\cal O}(r^0), \\
g_{++} &=   {\cal O} (r^0) ,
\end{aligned}
\label{Liouville_bndy_cond}
\end{equation}
where $F(x^+,x^-)$ satisfies $\partial_+\partial_-\log F=0$, yielding the boundary metric to have zero curvature. Therefore, the boundary conditions proposed by \cite{Troessaert:2013fma} fix the boundary metric to be flat up to a conformal factor; and further demanding that the boundary metric has vanishing Ricci curvature.\footnote{This corresponds to the $\chi=0$ case in section (\ref{liouville}), in (\ref{Liouvillesol}).} 
We now turn to generalising boundary conditions proposed by \cite{Troessaert:2013fma} to extended supergravity in $AdS_3$. For this we again use the CS formulation of $AdS_3$ gravity. The notations and conventions below are again taken from Henneaux $et\,\,al$ \cite{Henneaux:1999ib} and are summarised in Appendix \ref{Appendixextendedsugraconvention}. 

As the boundary conditions in this case are non-chiral the analyses for the two gauge fields are identical; hence we will give the details of only one of the gauge fields - $\tilde{A}$. One begins with an ansatz,
\bea
\tilde{A}&=&b\tilde{a}b^{-1}+bdb^{-1},\cr
\tilde{a}&=&\left[e^{-\tilde{\Phi}}\tilde{\kappa}\sigma^-+e^{\tilde{\Phi}}\sigma^+ + \tilde{B}_{a}T^a + \tilde{\psi}_{+\alpha}R^{+\alpha}+ \tilde{\psi}_{-\alpha}R^{-\alpha}\right]dx^-,\cr
\partial_+\tilde{a}&=&0,\,\,\,\,b=e^{\sigma^0\ln(r/\ell)}.
\eea  
The equation of motion, $\partial_+\tilde{a}_--\partial_-\tilde{a}_++[\tilde{a}_+,\tilde{a}_-]=0$ is readily satisfied. The above form of the gauge field ansatz doesn't need extra boundary terms added to the Chern-Simons action to make it variationally consistent. This is made apparent due to the fact that the gauge field $\tilde{a}$ as an 1-form only has a $dx^-$ component along the boundary while the fluctuation of the action yields,
\be
\delta S_{CS}[\tilde{A}]=\tfrac{k}{4\pi}\int_{\partial {\mathcal M}} Str[\tilde{a}\wedge\delta\tilde{a}].
\ee
 We now look for the space of gauge transformations which keep the above 1-form $\tilde{a}$ form-invariant.
\be
\delta_{\tilde{\Lambda}}\tilde{a}=d\tilde{\Lambda}+[\tilde{a},\tilde{\Lambda}]\,\,
\implies \delta_{\tilde{\Lambda}}\tilde{a}\big\vert_{\sigma^0}=0,\,\, \partial_+\tilde{\Lambda}=0
\ee
where 
\be
\tilde{\Lambda}=\tilde{\xi}_a\sigma^a+\tilde{b}_aT^a+\tilde{\epsilon}_{\pm\alpha}R^{\pm\alpha}.
\ee
Solving these constraints on $\tilde{\Lambda}$, we get
\bea
\delta_{\tilde{\Lambda}}\tilde{a}\big\vert_{\sigma^0}=0
\implies \tilde{\xi}_- = -\tfrac{1}{2}e^{-\tilde{\Phi}}\partial_-\tilde{\xi}_0+e^{-2\tilde{\Phi}}\tilde{\kappa}\tilde{\xi}_+-\tfrac{1}{2}e^{-\tilde{\Phi}}\eta^{\alpha\beta}(\tilde{\psi}_{+\alpha}\tilde{\epsilon}_{-\beta}+\tilde{\psi}_{-\alpha}\tilde{\epsilon}_{+\beta}).
\eea
The variations these gauge transformations induce on the functions parametrizing the 1-form $\tilde{a}$ are:
\bea
\delta_{\tilde{\Lambda}}\tilde{\Phi}&=&e^{-\tilde{\Phi}}\partial_-\tilde{\xi}_+
-\tilde{\xi}_0+i
e^{-\tilde{\Phi}}\eta^{\alpha\beta}\tilde{\psi}_{+\alpha}\tilde{\epsilon}_{+\beta},\cr
\delta_{\tilde{\Lambda}}\tilde{\kappa}&=&-\tfrac{1}{2}\partial^2_-\tilde{\xi}_0 + \tfrac{1}{2}\partial_-\tilde{\Phi}\partial_-\tilde{\xi}_0 -2\partial_-\tilde{\Phi}e^{-\tilde{\Phi}}\tilde{\kappa}\tilde{\xi}_+ e^{-\tilde{\Phi}}+\partial_-(\tilde{\kappa}\tilde{\xi}_+),\cr&&
-\tfrac{i}{2}\left[\partial_-\tilde{\Phi}-\partial_- \right]\eta^{\alpha\beta}(\tilde{\psi}_{+\alpha}\tilde{\epsilon}_{-\beta}+\tilde{\psi}_{-\alpha}\tilde{\epsilon}_{+\beta})+e^{-\tilde{\Phi}}\tilde{\kappa}\partial_-\tilde{\xi}_+\cr&& 
-i\eta^{\alpha\beta}\left[ e^{\tilde{\Phi}}\tilde{\psi}_{-\alpha}\tilde{\epsilon}_{-\beta}-e^{-\tilde{\Phi}}\tilde{\kappa}\tilde{\psi}_{+\alpha}\tilde{\epsilon_{+\beta}} \right],\cr
\delta_{\tilde{\Lambda}}\tilde{B}_{a}&=&\partial_-\tilde{b}_a+f_a^{\,\,bc}\tilde{B}_{b}\tilde{b}_c-i\tfrac{d-1}{2C_\rho}(\lambda^a)^{\alpha\beta}(\tilde{\psi}_{-\alpha}\tilde{\epsilon}_{+\beta}-\tilde{\psi}_{+\alpha}\tilde{\epsilon}_{-\beta}),\cr
\delta_{\tilde{\Lambda}}\tilde{\psi}_{+\alpha}&=&\partial_-\tilde{\epsilon}_{+\alpha}-\tfrac{1}{2}\tilde{\psi}_{+\alpha}\tilde{\xi}_0+(e^{\tilde{\Phi}}\tilde{\epsilon}_{-\alpha}-\tilde{\psi}_{-\alpha}\tilde{\xi}_+) - (\lambda^a)^\beta_{\,\,\alpha}(\tilde{B}_{a}\tilde{\epsilon}_{+\beta}-\tilde{b}_a\tilde{\psi}_{+\beta-}),\cr
\delta_{\tilde{\Lambda}}\tilde{\psi}_{-\alpha}&=&\partial_-\tilde{\epsilon}_{-\alpha}+\tfrac{1}{2}\tilde{\psi}_{-\alpha}\tilde{\xi}_0+e^{\tilde{\Phi}}\tilde{\kappa}\tilde{\epsilon}_{+\alpha}-(\lambda^a)^\beta_{\,\,\alpha}(\tilde{B}_{a}\tilde{\epsilon}_{-\beta}-\tilde{b}_a\tilde{\psi}_{-\beta}) \cr&&- \tilde{\psi}_{+\alpha}\left[ -\tfrac{1}{2}e^{-\tilde{\Phi}}\partial_-\tilde{\xi}_0+e^{-2\tilde{\Phi}}\tilde{\kappa}\tilde{\xi}_+ +\tfrac{i
}{2}e^{-\tilde{\Phi}}\eta^{\alpha\beta}(\tilde{\psi}_{+\alpha}\tilde{\epsilon}_{-\beta}-\tilde{\psi}_{-\alpha}\tilde{\epsilon}_{+\beta}) \right].
\eea
Associated to the above fluctuations are infinitesimal variations of a well defined asymptotic charge.
\bea
\mathrlap{\slash}\delta\tilde{Q}&=&-\tfrac{k}{4\pi}\int d\phi\,Str[\tilde{\Lambda}\delta\tilde{a}_\phi],\cr
&=&-\tfrac{k}{4\pi}\int d\phi\left\{ \tilde{\xi}_-\delta e^{\tilde{\Phi}}+\tilde{\xi}_+\delta(e^{-\tilde{\Phi}}\tilde{\kappa}) -i\eta^{\alpha\beta}(\delta\tilde{\psi}_{+\alpha}\tilde{\epsilon}_{-\beta}-\delta\tilde{\psi}_{-\alpha}\tilde{\epsilon}_{+\beta})+\tfrac{2C_\rho}{d-1}\delta\tilde{B}_{a}b^a \right\}.
\eea
The next step is to be able to write the above change in the charge such that it is a total variation $\delta \tilde{Q}$ so that $\delta$ can be taken out of the integral and (the charge) can be integrated from a suitable point (vacuum) on the solution space to any arbitrary point. Therefore, it is important to recognise field independent parameters parametrising the gauge transformations and accordingly functions of the fields parametrising the space of solutions on which a phase space can be defined via the Poisson brackets induced by the integrated charge.

The above expression for $\mathrlap{\slash}\delta\tilde{Q}$ can be simplified if one redefines 
\be
\tilde{\psi}_{\pm\alpha}=\tilde{\Psi}_{\pm\alpha}e^{\pm\tilde{\Phi}},\,\,\,\,\tilde{\epsilon}_{\pm\alpha}=\varepsilon_{\pm\alpha}e^{\pm\tilde{\Phi}},\,\,\,\,\tilde{\xi}_+=\tilde{\Xi}_+e^{\tilde{\Phi}}.
\label{redineTroesug}
\ee
\be
\delta\tilde{Q}=-\tfrac{k}{4\pi}\int d\phi\,\left\{ \tfrac{1}{2}\delta\tilde{\Phi}'\tilde{\xi}_0+\tilde{\Xi}_+\delta\tilde{\kappa}+\tfrac{2C_\rho}{d-1}\delta\tilde{B}_{a}\tilde{b}^a -i\eta^{\alpha\beta}\left[ \delta\tilde{\Psi}_{+\alpha}\tilde{\varepsilon}_{-\beta}-\delta\tilde{\Psi}_{-\alpha}\tilde{\varepsilon}_{+\beta} \right]\right\}
\ee  
Therefore the total integrated charge is
\be
\tilde{Q}=-\tfrac{k}{4\pi}\int d\phi\,\left\{ \tfrac{1}{2}\tilde{\Phi}'\tilde{\xi}_0+\tilde{\Xi}_+\tilde{\kappa}+\tfrac{2C_\rho}{d-1}\tilde{B}_{a}\tilde{b}^a -i\eta^{\alpha\beta}\left[ \tilde{\Psi}_{+\alpha}\tilde{\varepsilon}_{-\beta}-\tilde{\Psi}_{-\alpha}\tilde{\varepsilon}_{+\beta} \right]\right\}
\ee
The redefined gauge transformation parameters in (\ref{redineTroesug}) are therefore to be considered field independent. Also, one sees that it is $\tilde{\Phi}'$ and not $\tilde{\Phi}$ which is appropriate for defining the phase space structure on the space of solutions. The variations of the redefined fields in terms of the field independent gauge parameters are:
\bea
\delta_{\tilde{\Lambda}}\tilde{\Phi}'&=&\partial_-(\tilde{\Phi}'\tilde{\Xi}_+)+
\partial^2_-\tilde{\Xi}_+-\partial_-\tilde{\xi}_0+i\eta^{\alpha\beta}(\partial_-\tilde{\Psi}_{+\alpha}\tilde{\varepsilon}_{+\beta}+\tilde{\Psi}_{+\alpha}\partial_-\tilde{\varepsilon}_{+\beta}),\cr
\delta_{\tilde{\Lambda}}\tilde{\kappa}&=&2\tilde{\kappa}\partial_-\tilde{\Xi}_+
+\partial_-\tilde{\kappa}\tilde{\Xi}_+-\tfrac{1}{2}\partial_-^2\tilde{\xi}_0+\tfrac{1}{2}\tilde{\Phi}'\partial_-\tilde{\xi}_0 \cr
&&-\tfrac{i}{2}\eta^{\alpha\beta}\left[(\tilde{\Phi}'\tilde{\Psi}_{+\alpha}-\partial_-\tilde{\Psi}_{+\alpha}-\tilde{\Psi}_{+\alpha}\partial_-+2\tilde{\Psi}_{-\alpha})\tilde{\varepsilon}_{-\beta}  \right.\cr
&&\left.+(\tilde{\Phi}'\tilde{\Psi}_{-\alpha}-\partial_-\tilde{\Psi}_{-\alpha}-2\tilde{\kappa}\tilde{\Psi}_{+\alpha}-\tilde{\Psi}_{-\alpha}\partial_-)\tilde{\varepsilon}_{+\beta}  \right],\cr
\delta_{\tilde{\Lambda}}\tilde{\Psi}_{+\alpha}&=&-\tfrac{1}{2}\tilde{\Psi}_{+\alpha}\left[ \tilde{\Phi}'\tilde{\Xi}_++i\eta^{\rho\sigma}\tilde{\Psi}_{+\rho-}\tilde{\varepsilon}_{+\sigma} \right]+\partial_-\tilde{\varepsilon}_{+\alpha}+\tfrac{1}{2}\tilde{\Phi}'\tilde{\varepsilon}_{+\alpha}+\tilde{\varepsilon}_{-\alpha}-\tilde{\Psi}_{-\alpha}\tilde{\Xi}_+\cr&&-(\lambda^a)^\beta_{\,\,\alpha}(\tilde{B}_{a}\tilde{\varepsilon}_{+\beta}-\tilde{b}_a\tilde{\Psi}_{+\alpha}),\cr
\delta_{\tilde{\Lambda}}\tilde{\Psi}_{-\alpha}&=&\partial_-\tilde{\varepsilon}_{-\alpha} - \tfrac{1}{2}\tilde{\Phi}'\tilde{\varepsilon}_{-\alpha}+\tilde{\kappa}\tilde{\varepsilon}_{+\alpha} - (\lambda^a)^\beta_{\,\,\alpha}(\tilde{B}_{a}\tilde{\varepsilon}_{-\beta}-\tilde{b}_a\tilde{\Psi}_{-\beta-}) \cr&&- \tilde{\Psi}_{+\alpha}\left[ -\tfrac{1}{2}\partial_-\tilde{\xi}_0+\tilde{\kappa}\tilde{\Xi}_++i\eta^{\alpha\beta}(\tilde{\Psi}_{+\alpha}\tilde{\varepsilon}_{-\beta}-\tilde{\Psi}_{-\alpha}\tilde{\varepsilon}_{+\beta}) \right]\cr&&+\tfrac{1}{2}\tilde{\Psi}_{-\alpha}\left[ \tilde{\Psi}'\tilde{\Xi}_++\partial_-\tilde{\Xi}_++i\eta^{\alpha\beta}\tilde{\Psi}_{+\alpha}\tilde{\varepsilon}_{+\beta} \right],\cr
\delta_{\tilde{\Lambda}}\tilde{B}_{a}&=&\partial_-\tilde{b}_a+f_a^{\,\,bc}\tilde{B}_{a}\tilde{b}^a-i\tfrac{d-1}{2C\rho}(\lambda^a)^{\alpha\beta}(\tilde{\Psi}_{-\alpha}\tilde{\varepsilon}_{+\beta}-\tilde{\Psi}_{+\alpha}\tilde{\varepsilon}_{-\beta}).
\eea
This leads to the following Poisson brackets amongst the solution space variables
\bea
\tfrac{-k}{4\pi}\left\{ \tilde{\kappa}({x^-}'),\tilde{\kappa}(x^-) \right\}&=&-2\tilde{\kappa}\delta'({x^-}'-x^-)+\tilde{\kappa}'\delta({x^-}'-x^-),\cr
\tfrac{-k}{4\pi}\left\{ \tilde{\Phi}'({x^-}'),\tilde{\kappa}(x^-) \right\}&=&-\tilde{\Phi}'\delta'({x^-}'-x^-)-\delta''({x^-}'-x^-),\cr
\tfrac{-k}{4\pi}\left\{ \tilde{\Psi}_{+\alpha}({x^-}'),\tilde{\kappa}(x^-) \right\}&=&\left[ \tfrac{1}{2}\tilde{\Phi}'\tilde{\Psi}_{+\alpha}  -\tfrac{1}{2}\partial_-\tilde{\Psi}_{+\alpha}+\tilde{\Psi}_{-\alpha} \right]\delta({x^-}'-x^-)\cr&&+\tfrac{1}{2}\tilde{\Psi}_{+\alpha}(x^-)\delta'({x^-}'-x^-),\cr&&\cr
\tfrac{-k}{4\pi}\left\{ \tilde{\Psi}_{-\alpha}({x^-}'),\tilde{\kappa}(x^-) \right\}&=&\left[ -\tfrac{1}{2}\tilde{\Phi}'\tilde{\Psi}_{-\alpha}+\tfrac{1}{2}\partial_-\tilde{\Psi}_{-\alpha}-\tilde{\kappa}\tilde{\Psi}_{+\alpha} \right]\delta({x^-}'-x^-)\cr&&-\tfrac{1}{2}\tilde{\Psi}_{-\alpha}\delta'({x^-}'-x^-),\cr&&\cr
\tfrac{-k}{4\pi}\left\{ \tilde{\Phi}'({x^-}'),\tilde{\Phi}'(x^-) \right\}&=&2\delta'({x^-}'-x^-),\cr
\tfrac{-k}{4\pi}\left\{ \tilde{\Psi}_{-\alpha}({x^-}'),\tilde{\Phi}'(x^-) \right\}&=&-\tilde{\Psi}_{+\alpha}\delta'({x^-}'-x^-),\cr
\tfrac{-k}{4\pi}\left\{ \tilde{\Psi}_{+\alpha}({x^-}'),\tilde{\Psi}_{+\beta}(x^-) \right\}&=&-i\eta_{\alpha\beta}\delta({x^-}'-x^-),\cr
\tfrac{-k}{4\pi}\left\{ \tilde{\Psi}_{-\alpha}({x^-}'),\tilde{\Psi}_{+\beta}(x^-) \right\}&=&-i\eta_{\alpha\beta}\delta'({x^-}'-x^-)\cr
&&+\left[\tfrac{i}{2} \eta_{\alpha\beta}\tilde{\Phi}'-i(\lambda^a)_{\alpha\beta}\tilde{B}_{a}+\tfrac{1}{2}\tilde{\Psi}_{+\beta}\tilde{\Psi}_{+\alpha}\right]\delta({x^-}'-x^-),\cr
\tfrac{-k}{4\pi}\left\{ \tilde{\Psi}_{-\alpha}({x^-}'),\tilde{\Psi}_{-\beta}(x^-) \right\}&=&\left[ i\tilde{\kappa}\eta_{\alpha\beta}-\tilde{\Psi}_{+\beta}\tilde{\Psi}_{-\alpha}-\tfrac{1}{2}\tilde{\Psi}_{-\beta}\tilde{\Psi}_{+\alpha} \right]\delta({x^-}'-x^-),\cr
\tfrac{-k}{4\pi}\left\{ \tilde{B}_{a}({x^-}'),\tilde{\Psi}_{+\alpha}(x^-) \right\}&=&\tfrac{d-1}{2C\rho}(\lambda_a)^\beta_{\,\,\alpha}\tilde{\Psi}_{+\beta}\delta({x^-}'-x^-),\cr
\tfrac{-k}{4\pi}\left\{ \tilde{B}_{a}({x^-}'),\tilde{\Psi}_{-\alpha}(x^-) \right\}&=&\tfrac{d-1}{2C_\rho}(\lambda_a)^\beta_{\,\,\alpha}\tilde{\Psi}_{-\beta}\delta({x^-}'-x^-),\cr
\tfrac{-k}{4\pi}\left\{ \tilde{B}_{a}({x^-}'),\tilde{B}_{b}(x^-) \right\}&=&-\tfrac{d-1}{2C_\rho}\left[ \delta_{ab}\delta'({x^-}'-x^-)-f_{ab}^{\,\,\,\,\,c}\tilde{B}_{c}\delta({x^-}'-x^-) \right].
\eea
The first Poisson bracket among $\tilde{\kappa}$ s can be easily recognised as that of the Witt algebra (Virasoro without the central extension). As was done in \cite{Troessaert:2013fma} (and advocated against in \cite{Apolo:2014tua}) the phase-space variable $\tilde{\Phi}'$ can be used to generate the central term in the Witt algebra of $\tilde{\kappa}$ by redefining 
\be
\hat{\tilde{\kappa}}=\tilde{\kappa}+\alpha_c \tilde{\Phi}''.
\ee  
After further rescaling $\hat{\tilde{\kappa}}\rightarrow\tfrac{k}{2\pi}$ one can show the central term in the Virasoro of $\hat{\tilde{\kappa}}$ to be $2\alpha_c(\alpha_c+1)k/(2\pi)$; which for $\alpha_c=-\tfrac{1}{2}\pm\tfrac{1}{\sqrt{2}}$ yields the central extension of the Virasoro found in Brown-Henneaux analysis.
One can mode expand the above Poisson brackets and find the  relevant algebra. Depending on integer or half-integer moding of the fermionic charges the algebra will either fall into Ramond or the NS sector respectively.

\section{Flat limit of extended $AdS_3$ supergravity}
\label{Flat_sugra}
In this section we deviate from the main theme of the paper so far to suggest a possible flat limit of extended $AdS_3$ supergravity studied by Henneaux $et\,al$ \cite{Henneaux:1999ib}. In \cite{Henneaux:1999ib} a thorough asymptotic symmetry analysis of all allowed but arbitrary $AdS_3$ supergravities was carried out. The boundary conditions on these $AdS_3$ supergravities are the most general ones which are consistent with the boundary conditions proposed by Brown and Henneaux for $AdS_3$ gravity. The boundary metric in \cite{Henneaux:1999ib} is that of 2d flat Minkowski. The asymptotic symmetries in the case of such generic $AdS_3$ supergravities was found to be a super extension of the Virasoro algebra with quadratic non-linearities\footnote{These quadratic non-linearities are non-trivial in that they cannot be redefined away.}.

In this section we answer the question whether similar boundary conditions exist in flat 3d Minkowski space and what could be their asymptotic symmetry algebra. We answer this by taking a suitable flat space limit-$l\rightarrow\infty$; of the analysis in \cite{Henneaux:1999ib}. Since the analysis in \cite{Henneaux:1999ib} maintains the boundary metric to be of 2d Minkowski, the $l\rightarrow\infty$ limit would correspond limits of such supergravities with a flat boundary metric in $AdS$. 

Such limits were first considered by \cite{Barnich:2012aw} and subsequently investigated in \cite{Barnich:2014cwa,Banerjee:2016nio,Banerjee:2017gzj,Fuentealba:2017fck}. Here the specific $AdS_3$ supergravities were considered which did not yield any non-linearities in their asymptotic symmetry algebra. In this section we show that certain flat limits exists which allow us to retain the non-linearities found in \cite{Henneaux:1999ib}. We begin by first analysing different $l\rightarrow\infty$ limits of  the graded algebras considered in \cite{Henneaux:1999ib}. In principle one can then take any of these limits on the solutions of $AdS_3$ supergravity in \cite{Henneaux:1999ib} and therefore proceed analogously to determine boundary conditions, compute asymptotic charges and determine the asymptotic symmetry algebra \cite{Barnich:2014cwa,Banerjee:2016nio,Banerjee:2017gzj,Fuentealba:2017fck}. We however choose one out these many possible limits which enable us to retain the quadratic non-linearities even in the flat space.
\\\\  
The $SL(2,{\mathbb R})$ algebra we use here differ from that in \cite{Henneaux:1999ib} as $L_{\pm}=\pm \sigma^{\mp}$ and $L_0=\sigma^0$. Therefore the gauge fields of \cite{Henneaux:1999ib} after having their $r$-dependence stripped off take the form
\bea
a^+&=&\left[L_+-\kappa(x^+)L_-+B_a(x^+)T^a+Q_{+\alpha}(x^+)R^{+\alpha}\right]dx^+\cr
a^-&=&\left[-\bar{L}_-+\bar{\kappa}(x^-)\bar{L}_++\bar{B}_a(x^-)\bar{T}^a+\bar{Q}_{-\alpha}(x^-)\bar{R}^{-\alpha}\right]dx^-.
\label{boundary_super_gauge_filed}
\eea 
We then perform the finite gauge transformation given by Barnich {\it et al} \cite{Barnich:2014cwa, Barnich:2012aw}:
\bea
A^+&=&g_+^{-1} \, a \, g_++g_+^{-1} \, dg_+,\cr
A^-&=&g_-^{-1} \, a \, g_-+g_-^{-1}\, dg_-,\cr\cr
{\rm where}\,\,\,\, g_+&=&e^{(r/2l)L_-}, ~~~~ g_-=e^{-\log(r^2/4l^2)\bar{L}_0}  ~ e^{(r/2l)\bar{L}_-} ~ e^{(2l/r)\bar{L}_+}
\eea
Note the asymmetric manner in which the two gauge fields are being treated under the finite gauge transformation. One can proceed in an even handed manner but this doesn't yield the known metric in the $l\rightarrow\infty$ limit at ${\mathcal{I}}^+$ in BMS gauge\footnote{It may be related to the asymptotically flat metric at ${\mathcal{I}}^+$ by a large gauge transformation which mix the null and the boundary co-ordinates, thus introducing changes in charges; $i.e.$ it might not be a small gauge transformation in the $l\rightarrow\infty$ limit. }. After the gauge transformations the gauge fields take the form
\bea
A^+&=&\frac{L_-}{2l}dr+\left[\frac{r}{l}L_0+L_++L_-\left(\frac{r^2}{4l^2}-\kappa(x^+)\right)+B^+_a(x^+)T^{+a}+Q_{+\alpha}(x^+)R^{+\alpha}\right]dx^+,\cr
A^-&=&\frac{-\bar{L}_-}{2l}dr+\left[\frac{r}{l}\bar{L}_0-\bar{L}_++\bar{L}_-\left(-\frac{r^2}{4l^2}+\bar{\kappa}(x^-)\right)+B^-_a(x^-)T^{-a}+\bar{Q}_{-\alpha}(x^-)\bar{R}^{+\alpha}\right]dx^-.  \nonumber \\
\eea
The solution to the flat space equation of motion would be given by the sum $A=A^++A^-$ in the limit $l\rightarrow\infty$ after replacing $x^\pm=\frac{u}{l}\pm\phi$.

\subsection{$l\rightarrow\infty$ limit of the algebra}
We redefine (such as in \cite{Bagchi:2009my}) the spacetime algebra elements as follows:
\bea
2L_0=J_2+lP_2,&&2\bar{L}_0=J_2-lP_2,\cr
2L_+=J_1+lP_1,&&2\bar{L}_+=J_1-lP_1,\cr
-L_-=J_0+lP_0,&&-\bar{L}_-=J_0-lP_0.
\eea
Thus yielding
\bea
&&[J_0,J_2]=-J_0,\hspace{8mm}[J_0,J_1]=J_2,\hspace{9mm}[J_1,J_2]=J_1,\cr
&&[J_0,P_2]=-P_0,\hspace{7mm}[J_0,P_1]=P_2,\hspace{9mm}[J_1,P_2]=P_1,\cr
&&[P_0,J_2]=-P_0,\hspace{7mm}[P_0,J_1]=P_2,\hspace{9mm}[P_1,J_2]=P_1,
\label{iso12}
\eea
where $P_i$ are translations that commute among themselves only in $l\rightarrow\infty$. We further choose to define
\bea
f(l)S^{\pm \alpha}=(R^{\pm\alpha}+\bar{R}^{\pm\alpha}), && \bar{f}(l)\bar{S}^{\pm \alpha}=(R^{\pm\alpha}-\bar{R}^{\pm\alpha}),\cr
g(l)T_a=(T_a^++T_a^-), && \bar{g}(l)\bar{T}_a=(T_a^+-T_a^-).
\eea
here the functions $f(l),\bar{f}(l),g(l)$ \& $\bar{g}(l)$ are as of yet undefined.The commutators independent of these functions are
\bea
[J_2,S^{\pm\alpha}]=\tfrac{\pm}{2}S^{\pm\alpha},&&[J_2,\bar{S}^{\pm\alpha}]=\pm\tfrac{1}{2}\bar{S}^{\pm\alpha},\cr
[J_1,S^{+\alpha}]=S^{-\alpha},&&[J_1,\bar{S}^{+\alpha}]=\bar{S}^{-\alpha},\cr
[J_0,S^{-\alpha}]=\tfrac{1}{2}S^{+\alpha},&&[J_0,\bar{S}^{-\alpha}]=\tfrac{1}{2}\bar{S}^{+\alpha}.
\label{iso12super}
\eea
While the ones which do depend on them are
\bea
lf(l)[P_2,S^{\pm\alpha}]=\pm\tfrac{1}{2}\bar{f}(l)\bar{S}^{\pm\alpha},&& l\bar{f}(l)[P_2,\bar{S}^{\pm\alpha}]=\pm\tfrac{1}{2}f(l)S^{\pm\alpha},\cr
lf(l)[P_1,S^{+\alpha}]=\bar{f}(l)\bar{S}^{-\alpha},&& l\bar{f}(l)[P_1,\bar{S}^{+\alpha}]=f(l)S^{-\alpha},\cr
lf(l)[P_0,S^{-\alpha}]=\tfrac{1}{2}\bar{f}(l)\bar{S}^{+\alpha},&& l\bar{f}(l)[P_0,\bar{S}^{-\alpha}]=\tfrac{1}{2}f(l)S^{+\alpha}.
\eea
\bea
f(l)^2\{S^{+\alpha},S^{+\beta}\}=2\eta^{\alpha\beta}J_0,&&\bar{f}(l)^2\{\bar{S}^{+\alpha},\bar{S}^{+\beta}\}=2\eta^{\alpha\beta}J_0,\cr
f(l)^2\{S^{-\alpha},S^{-\beta}\}=-\eta^{\alpha\beta}J_1,&&\bar{f}(l)^2\{\bar{S}^{-\alpha},\bar{S}^{-\beta}\}=-\eta^{\alpha\beta}J_1,
\label{SStoJ}
\eea
\bea
f(l)\bar{f}(l)\{S^{+\alpha},\bar{S}^{+\beta}\}=2\eta^{\alpha\beta}lP_0,&&
f(l)\bar{f}(l)\{S^{-\alpha},\bar{S}^{-\beta}\}=-\eta^{\alpha\beta}lP_1,
\label{SStoP}
\eea
and
\bea
f(l)^2\{S^{\pm\alpha},S^{\mp\beta}\}&=&-\eta^{\alpha\beta}J_2\pm
(\tfrac{d-1}{2C_\rho})\Lambda_a^{\alpha\beta}g(l)T_a,\cr\bar{f}(l)^2\{\bar{S}^{\pm\alpha},\bar{S}^{\mp\beta}\}&=&-\eta^{\alpha\beta}J_2\pm
(\tfrac{d-1}{2C_\rho})\Lambda_a^{\alpha\beta}g(l)T_a,\cr
f(l)\bar{f}(l)\{S^{\pm\alpha},\bar{S}^{\mp\beta}\}&=&-\eta^{\alpha\beta}lP_2\pm
(\tfrac{d-1}{2C_\rho})\Lambda_a^{\alpha\beta}\bar{g}(l)\bar{T}_a.
\label{SStoJandT}
\eea
This is the term which is indeed responsible for the quadratic terms in the r.h.s. in the asymptotic symmetry algebra of \cite{Henneaux:1999ib}. The internal bosonic symmetries scale like
\be
g(l)[T_a,T_b]=f_{abc}T_c,\,\,\,\,\,\bar{g}(l)^2[\bar{T}_a,\bar{T}_b]=f_{abc}g(l)T_c,\,\,\,\,\,g(l)[T_a,\bar{T}_b]=f_{abc}\bar{T}_c
\label{TT}
\ee
Further,
\bea
[T_a,S^{\pm\alpha}]=-\tfrac{1}{g}(\Lambda_a)^\alpha_\beta S^{\pm\beta},&& [\bar{T}_a,S^{\pm\alpha}]=-\left(\frac{\bar{f}}{f\bar{g}}\right)(\Lambda_a)^\alpha_\beta \bar{S}^{\pm\beta},\cr
[T_a,\bar{S}^{\pm\alpha}]=-\tfrac{1}{g}(\Lambda_a)^\alpha_\beta \bar{S}^{\pm\beta},&& [\bar{T}_a,\bar{S}^{\pm\alpha}]=-\left(\frac{f}{\bar{f}\bar{g}}\right)(\Lambda_a)^\alpha_\beta S^{\pm\beta}.
\eea
\newline
Now we would like to specify the functions $f(l),\bar{f}(l),g(l), \bar{g}(l)$ as powers in $l$ such that a consistent $l\rightarrow\infty$ limit exists for the algebra $i.e.$ the $r.h.s.$ of the above algebra should not diverge as $l\rightarrow\infty$.
\vskip .5cm
\noindent\underline{\bf Different limits as $l\rightarrow\infty$}
\vskip .5cm
%
We start by observing that (\ref{SStoP}) demands $f\bar{f}\geq l$ and $2^{nd}$ eq. of (\ref{TT}) demands $\bar{g}^2\geq g$. We will categorize them as follows with (\ref{iso12}) and (\ref{iso12super}) holding in all cases below:
\subsubsection*{CaseI:\,\,$f=\bar{f}$}
\underline{1). $f=\bar{f}=\sqrt{l}$ and $g=\bar{g}=l$}:
\bea
\{S^{+\alpha},\bar{S}^{+\beta}\}=-2\eta^{\alpha\beta}P_0,&&
\{S^{-\alpha},\bar{S}^{-\beta}\}=\eta^{\alpha\beta}P_1,
\cr
\{S^{+\alpha},S^{-\beta}\}=
\left(\tfrac{d-1}{2C_\rho}\right)\Lambda_a^{\alpha\beta}T_a,&&\{\bar{S}^{+\alpha},\bar{S}^{-\beta}\}=\left(\tfrac{d-1}{2C_\rho}\right)\Lambda_a^{\alpha\beta}T_a,\cr
\{S^{+\alpha},\bar{S}^{-\beta}\}=-\eta^{\alpha\beta}P_2+
\left(\tfrac{d-1}{2C_\rho}\right)\Lambda_a^{\alpha\beta}\bar{T}_a,
&&\{\bar{S}^{+\alpha},S^{-\beta}\}=-\eta^{\alpha\beta}P_2+
\left(\tfrac{d-1}{2C_\rho}\right)\Lambda_a^{\alpha\beta}\bar{T}_a,\cr
[T_a,T_b]=0,\,\,\,&&[\bar{T}_a,\bar{T}_b]=0,\,\,\,\,\,[T_a,\bar{T}_b]=0.
\label{caseI1}
\eea
\newline\newline
\underline{2). $f=\bar{f}=\sqrt{l}$ and $g=\bar{g}=\sqrt{l}$}:
\bea
\{S^{+\alpha},\bar{S}^{+\beta}\}=-2\eta^{\alpha\beta}P_0,&&
\{S^{-\alpha},\bar{S}^{-\beta}\}=\eta^{\alpha\beta}P_1,
\cr
\{S^{+\alpha},S^{-\beta}\}=0,&&\{\bar{S}^{+\alpha},\bar{S}^{-\beta}\}=0,\cr
\{S^{+\alpha},\bar{S}^{-\beta}\}=-\eta^{\alpha\beta}P_2,&&\{\bar{S}^{+\alpha},S^{-\beta}\}=-\eta^{\alpha\beta}P_2,\cr
[T_a,T_b]=0,\,\,\,&&[\bar{T}_a,\bar{T}_b]=0,\,\,\,\,\,[T_a,\bar{T}_b]=0.
\label{caseI2}
\eea
\underline{3). $f=\bar{f}=\sqrt{l}$ and $g=1,\,\,\bar{g}=l$}:
\bea
\{S^{+\alpha},\bar{S}^{+\beta}\}=-2\eta^{\alpha\beta}P_0,&&
\{S^{-\alpha},\bar{S}^{-\beta}\}=\eta^{\alpha\beta}P_1,
\cr
\{S^{+\alpha},S^{-\beta}\}=0,&&\{\bar{S}^{+\alpha},\bar{S}^{-\beta}\}=0,\cr
\{S^{+\alpha},\bar{S}^{-\beta}\}=-\eta^{\alpha\beta}P_2+
\left(\tfrac{d-1}{2C_\rho}\right)\Lambda_a^{\alpha\beta}\bar{T}_a,
&&\{\bar{S}^{+\alpha},S^{-\beta}\}=-\eta^{\alpha\beta}P_2+
\left(\tfrac{d-1}{2C_\rho}\right)\Lambda_a^{\alpha\beta}\bar{T}_a,\cr
[T_a,T_b]=f_{abc}T_c,\,\,\,\,\,&&[\bar{T}_a,\bar{T}_b]=0,\,\,\,\,\,[T_a,\bar{T}_b]=f_{abc}\bar{T}_c.\cr
[T_a,S^{\pm\alpha}]=-(\Lambda_a)^\alpha_\beta S^{\pm\beta}&&[T_a,\bar{S}^{\pm\alpha}]=-(\Lambda_a)^\alpha_\beta \bar{S}^{\pm\beta}
\label{caseI3}
\eea
The algebra studied in \cite{Banerjee:2016nio,Banerjee:2017gzj,Fuentealba:2017fck} fall in this last category. Notice in the all above cases the anti-commutators never produce a $J_i$ since both $f$ \& $\bar{f}$ scale with $l$. This will not be true in the next case.
\subsubsection*{Case II\,: \, \underline{$f=l,\,\,\bar{f}=1$ and $g=1,\,\,\bar{g}=l$} ~:}
\bea
[P_2,S^{\pm\alpha}]=0,&& [P_2,\bar{S}^{\pm\alpha}]=\pm\tfrac{1}{2}S^{\pm\alpha},\cr
[P_1,S^{+\alpha}]=0,&& [P_1,\bar{S}^{+\alpha}]=S^{-\alpha},\cr
[P_0,S^{-\alpha}]=0,&& [P_0,\bar{S}^{-\alpha}]=\tfrac{1}{2}S^{+\alpha}.
\nonumber
\eea
\bea
\{S^{+\alpha},S^{+\beta}\}=0,&&\{\bar{S}^{+\alpha},\bar{S}^{+\beta}\}=2\eta^{\alpha\beta}J_0,\cr
\{S^{-\alpha},S^{-\beta}\}=0,&&\{\bar{S}^{-\alpha},\bar{S}^{-\beta}\}=-\eta^{\alpha\beta}J_1,\cr
\{S^{+\alpha},\bar{S}^{+\beta}\}=2\eta^{\alpha\beta}P_0,&&
\{S^{-\alpha},\bar{S}^{-\beta}\}=-\eta^{\alpha\beta}P_1,
\cr
\{S^{\pm\alpha},S^{\mp\beta}\}=0,&&\{\bar{S}^{\pm\alpha},\bar{S}^{\mp\beta}\}=-\eta^{\alpha\beta}J_2\pm
(\tfrac{d-1}{2C_\rho})\Lambda_a^{\alpha\beta}T_a,\cr
\{S^{\pm\alpha},\bar{S}^{\mp\beta}\}=-\eta^{\alpha\beta}P_2\pm
(\tfrac{d-1}{2C_\rho})\Lambda_a^{\alpha\beta}\bar{T}_a,
&&\{\bar{S}^{\pm\alpha},S^{\mp\beta}\}=-\eta^{\alpha\beta}P_2\pm
(\tfrac{d-1}{2C_\rho})\Lambda_a^{\alpha\beta}\bar{T}_a,\cr
&&\cr
[T_a,T_b]=f_{abc}T_c,\,\,\,\,\,&&[\bar{T}_a,\bar{T}_b]=0,\,\,\,\,\,[T_a,\bar{T}_b]=f_{abc}\bar{T}_c,\cr
&&\cr
[T_a,S^{\pm\alpha}]=-(\Lambda_a)^\alpha_\beta S^{\pm\beta},&& [\bar{T}_a,S^{\pm\alpha}]=0,\cr
[T_a,\bar{S}^{\pm\alpha}]=-(\Lambda_a)^\alpha_\beta \bar{S}^{\pm\beta},&& [\bar{T}_a,\bar{S}^{\pm\alpha}]=-(\Lambda_a)^\alpha_\beta S^{\pm\beta}.
\label{caseII}
\eea

\subsection{Charges and Symmetry algebras}
We now proceed to express $A=A^+ + A^-$ for the different cases mentioned above.
\bea
A&=&A^+ + A^-\cr
&=& P_0\left(-dr-\frac{r^2}{2l^2}du + Mdu+Nd\phi \right)+P_1du+rP_2d\phi\cr
&&\,\,+J_0\left(-\frac{r^2}{2l^2}d\phi + \frac{N}{l^2}du+Md\phi \right)+J_1d\phi+\frac{r}{l^2}J_2du\cr
&&\,\,+\left[\frac{du}{2l}((B^+_a+B^-_a)gT_a+(B^+_a-B^-_a)\bar{g}\bar{T}_a)+\frac{d\phi}{2}((B^+_a-B^-_a)gT_a+(B^+_a+B^-_a)\bar{g}\bar{T}_a)\right]\cr
&&\,\,+\left[\frac{du}{2l}((Q^{+\alpha}+\bar{Q}_{-\alpha})fS^{+\alpha}
+(Q^{+\alpha}-\bar{Q}_{-\alpha})\bar{f}\bar{S}^{+\alpha})\right.\cr
&&\hspace{0.8cm}\left.+\frac{d\phi}{2}((Q^{+\alpha}-\bar{Q}_{-\alpha})fS^{+\alpha}
+(Q^{+\alpha}+\bar{Q}_{-\alpha})\bar{f}\bar{S}^{+\alpha})\right],
\label{genricflatA}
\eea
where $M=\kappa+\bar{\kappa},\,N=l(\kappa-\bar{\kappa})$ and $x^\pm=\tfrac{u}{l}\pm\phi$. We have furter supressed the coordinate dependencies for simplicity. From here on one would have to make a consistent choice of functions $\{f,\bar{f},g,\bar{g}\}$ as specified in the last subsection.
\newline\newline 
We will workout the limits in the last case above, namely, Case II: $f=\bar{g}=l \,\,\&\, \bar{f}=g=1$.
In order that the terms in \eqref{genricflatA} do not blow-up in the $l \rightarrow \infty$ limit we redefine
\bea
l(B_a^+ + B_a^-)= B_a && (B_a^+ - B_a^-)= \bar{B}_a,\cr
(Q_{+\alpha}+\bar{Q}_{-\alpha})= Q_{\alpha} && l(Q_{+\alpha}-\bar{Q}_{-\alpha})= \bar{Q}_{\alpha}.
\eea
recalling that $\kappa, B_a^+$ and $Q_{+\alpha} $ only depended on $x^+$ while $\bar{\kappa},B_a^- $ and $\bar{Q}_{-\alpha} $ on $x^-$, and given that $x^\pm=\tfrac{u}{l}\pm\phi$ we get
\be
\partial_u M=\frac{1}{l^2}\partial_\phi N\overset{l\rightarrow\infty}{=}0,\hspace{0.8cm}\partial_u\bar{B}_a=\frac{1}{l^2}\partial_\phi B_a\overset{l\rightarrow\infty}{=}0,\hspace{0.8cm}\partial_u Q_\alpha = \frac{1}{l^2}\partial_\phi\bar{Q}_\alpha\overset{l\rightarrow\infty}{=}0
\ee
The gauge field \eqref{genricflatA} after taking the limit $l\rightarrow\infty$ looks like
\bea
A&=&P_0(Mdu+Nd\phi) +P_1du +rP_2d\phi+J_0 Md\phi +J_1d\phi-P_0dr\cr
&&+\tfrac{1}{2}\bar{T^a}\bar{B}_a du +\tfrac{1}{2}(T^a\bar{B}_a+\bar{T}^aB_a)d\phi\cr
&&+\tfrac{1}{l}Q_\alpha S^{+\alpha}du+\tfrac{1}{2}(\bar{Q}_\alpha S^{+\alpha}+Q_\alpha\bar{S}^{+\alpha})d\phi.
\eea 
The $r$-dependence can be further gauged away by working with
\bea
a&=&e^{-rP_0}A \, e^{rP_0}+P_0dr\cr
&=&P_0(Mdu+Nd\phi) +P_1du +J_0 Md\phi +J_1d\phi\cr
&&+\tfrac{1}{2}\bar{T^a}\bar{B}_a du +\tfrac{1}{2}(T^a\bar{B}_a+\bar{T}^aB_a)d\phi\cr
&&+\tfrac{1}{l}Q_\alpha S^{+\alpha}du+\tfrac{1}{2}(\bar{Q}_\alpha S^{+\alpha}+Q_\alpha\bar{S}^{+\alpha})d\phi.
\eea
where the $l\rightarrow\infty$ limit implies
\be
\partial_uN=\partial_\phi M,\hspace{0.5cm}\partial_u B_a=\partial_\phi \bar{B}_a,\hspace{0.5cm}\partial_u \bar{Q}_\alpha=\partial_\phi Q_\alpha.
\ee
The infinitesimal gauge transformation that keep the above gauge field form-invariant is
\bea
\lambda&=&\xi^iP_i+\chi^iJ_i+b_aT^a+\bar{b}_a\bar{T}^a
+\psi_{\pm\alpha}S^{\pm\alpha}
+\bar{\psi}_{\pm\alpha}\bar{S}^{\pm\alpha},\cr&&\cr
\xi^2&=&-\partial_\phi\xi^1,\hspace{0.5cm}\chi^2=-\partial_\phi \chi^1=-\partial_u\xi^1,\cr
\partial_u\chi^1&=&0,\hspace{0.5cm}\partial_u b^a=0,\hspace{.5cm}\partial_u\bar{\psi}_{\pm\alpha}=0,\cr
\partial_u\psi_{-\alpha}&=&\partial_\phi\bar{\psi}_{-\alpha},\hspace{0.5cm}\partial_u\bar{b}_a=\partial_\phi b_a,\cr
\xi^0&=&-\partial^2_\phi\xi^1+N\chi^1+M\xi^1-\tfrac{i}{2}\bar{Q}_\alpha\bar{\psi}_{-\beta}\eta^{\alpha\beta}-\tfrac{i}{2}Q_\alpha\psi_{-\beta}\eta^{\alpha\beta},\cr
\chi^0&=-&\partial^2_\phi\chi^1+M\chi^1-\tfrac{i}{2}Q_\alpha\bar{\psi}_{-\beta}\eta^{\alpha\beta},\cr
\psi_{+\alpha}&=&-\partial_\phi\psi_{-\alpha}+\tfrac{1}{2}\bar{Q}_\alpha\chi^1+\tfrac{1}{2}Q_\alpha\xi^1+\tfrac{1}{2}\bar{B}_a\psi_{-\beta}(\Lambda^a)^\beta_\alpha+\tfrac{1}{2}B_a\bar{\psi}_{-\beta}(\Lambda^a)^\beta_\alpha,\cr
\bar{\psi}_{+\alpha}&=&-\partial_\phi\bar{\psi}_{-\alpha}+\tfrac{1}{2}\bar{B}_a\bar{\psi}_{-\beta}(\Lambda^a)^\beta_\alpha+\tfrac{1}{2}Q_\alpha\chi^1.
\eea
The corresponding fluctuations are
\bea
\delta M&=&\partial_\phi \chi^0-M\chi^2+iQ_\alpha\bar{\psi}^{+\alpha},\cr
\delta N&=&\partial_\phi \xi^0 -N\chi^2-M\xi^2+i\bar{Q}_\alpha\bar{\psi}^{+\alpha}+iQ_\alpha\psi^{+\alpha},\cr
\delta\bar{B}_a&=&2\partial_\phi b_a+f_a^{\,\,bc}\bar{B}_bb_c+iQ_\alpha\bar{\psi}_{-\beta}(\Lambda_a)^{\alpha\beta}(\tfrac{d-1}{2C_\rho}),\cr
\delta B_a&=&2\partial_\phi\bar{b}_a+f_a^{\,\,bc}(\bar{B}_b\bar{b}_c+B_bb_c)+i(Q_\alpha\psi_{-\beta}+\bar{Q}_\alpha \bar{\psi}_{-\beta}),\cr
\delta Q_\alpha&=&2\partial_\phi\bar{\psi}_{+\alpha}-\tfrac{1}{2}Q_\alpha\chi^2 + M\bar{\psi}_{-\alpha}+ (Q_\beta b_a -\bar{B}_a\bar{\psi}_{+\beta})(\Lambda^a)^\beta_\alpha,\cr
\delta \bar{Q}_\alpha &=& 2\partial_\phi \psi_{+\alpha}-\tfrac{1}{2}(\bar{Q}_\alpha\chi^2 + Q_\alpha\xi^2) + M\psi_{-\alpha} + N\bar{\psi}_{-\alpha} \cr
&&\hspace{0.1cm}+(\Lambda^a)^\beta_\alpha(Q_\beta\bar{b}_a + \bar{Q}_\beta b_a - \bar{B}_a\psi_{+\beta} - B_a\bar{\psi}_{+\beta}).   
\eea
In order to express the difference of two CS theories- each valued in $sl(2,\mathbb{R})$; as one CS theory valued in $sl(2,\mathbb{R})\times sl(2,\mathbb{R})$, the (super) trace (Killing metric) in one must differ from the other in sign. Therefore the non-zero Killing metric components are:	
\bea
&&\langle P_0,J_1\rangle=\langle P_1,J_0\rangle=\langle P_2,J_2\rangle=\frac{1}{l},\cr
&&\langle S^{-\alpha},\bar{S}^{+\beta}\rangle=\langle \bar{S}^{-\alpha},S^{+\beta}\rangle=\tfrac{2}{l}\eta^{\alpha\beta}	=-\langle S^{+\alpha},\bar{S}^{-\beta}\rangle=-\langle \bar{S}^{+\alpha},S^{-\beta}\rangle,\cr
&&\langle T_a,\bar{T}_b\rangle=\frac{4C_\rho}{(d-1)l}\delta_{ab}.\hspace*{1cm}({\rm where}\,f=l=\bar{g},\bar{f}=g=1)
\eea  
The $\frac{1}{l}$ in the super trace exactly cancels the $l$ in front of the action integral. The symplectic structure is then given by
\bea
\mathrlap{\slash}\delta Q_\lambda&=&\frac{1}{16\pi G}\oint d\phi\,\langle \delta a_\phi,\lambda \rangle\cr
&=&\frac{1}{16\pi G}\oint d\phi\, \{ \chi^1\delta N_{(0)}+\xi^1_{(0)}\delta M +\left(\tfrac{2C_\rho}{d-1} \right)(\bar{b}^a_{(0)}\delta\bar{B}_a + b_a\delta B^a_{(0)})\cr
&&\hspace{2.4cm}+i\bar{\psi}_{-\alpha}\delta\bar{Q}^\alpha_{(0)} +i \psi^{-\alpha}_{(0)}\delta Q_\alpha\}.
\eea
Integrating on the space of fluctuations we get
\bea
Q&=&\frac{1}{16\pi G}\oint d\phi\, \{ \chi^1 N_{(0)}+\xi^1_{(0)} M +\left(\tfrac{2C_\rho}{(d-1)} \right)(\bar{b}^a_{(0)}\bar{B}_a + b_a B^a_{(0)})\cr
&&\hspace{2.4cm}+i\bar{\psi}_{-\alpha}\bar{Q}^\alpha_{(0)} +i \psi^{-\alpha}_{(0)} Q_\alpha\}.
\eea
where we have used function which only depends on $\phi$, defined as;
\bea
&&N=N_{(0)}+u\partial_\phi M,\hspace{0.5cm}B_a=B_{a(0)}+u\partial_\phi\bar{B}_a,\hspace{0.5cm}\bar{Q}_\alpha=\bar{Q}_{\alpha(0)}+u\partial_\phi Q_\alpha,\cr
&&\xi^1=\xi^1_{(0)}+u\partial_\phi \chi^1,\hspace{0.5cm}\bar{b}_a=\bar{b}_{a(0)}+u\partial_\phi b_a,\hspace{0.5cm}\psi_{-\alpha}=\psi_{-\alpha(0)} + u\partial_\phi\bar{\psi}_{-\alpha}.
\eea
Evaluating the Poisson brackets one finds:
\bea
\tfrac{-1}{16\pi G}\left\{M(\phi),N^{(0)}(\phi')\right\}&=&-\delta'''(\phi-\phi')+(M(\phi)+M(\phi'))\delta'(\phi-\phi'),\cr
\tfrac{-1}{16\pi G}\left\{M(\phi),\bar{Q}^{(0)}_\alpha(\phi')\right\}&=&-\tfrac{1}{2}\bar{B}_a(\phi)Q_\beta(\phi)(\Lambda^a)_\alpha^{\,\,\,\beta}\delta(\phi-\phi') + \left(\tfrac{1}{2}Q_\alpha(\phi')+Q_\alpha(\phi)\right)\delta'(\phi-\phi'),\cr
\tfrac{-1}{16\pi G}\left\{N^{(0)}(\phi),N^{(0)}(\phi')\right\}&=&
\left(N_{(0)}(\phi')+N_{(0)}(\phi)\right)\delta'(\phi-\phi'),\cr
\tfrac{-1}{16\pi G}\left\{N^{(0)}(\phi),Q_\alpha(\phi')\right\}&=&-\tfrac{1}{2}\bar{B}_a(\phi)Q_\beta(\phi)(\Lambda^a)_\alpha^{\,\,\,\beta}\delta(\phi-\phi')+\left(\tfrac{1}{2}Q_\alpha(\phi')+Q_\alpha(\phi)\right)\delta'(\phi-\phi'),\cr
\tfrac{-1}{16\pi G}\left\{N^{(0)}(\phi),\bar{Q}^{(0)}_\alpha(\phi')\right\}&=&-\tfrac{1}{2}(B^{(0)}_a(\phi)Q_\beta(\phi)+\bar{B}_a(\phi)\bar{Q}^{(0)}_\beta(\phi))(\Lambda^a)_\alpha^{\,\,\,\beta}\delta(\phi-\phi')\cr
&&\hspace{2cm}+\left(\tfrac{1}{2}\bar{Q}^{(0)}_\alpha(\phi')+\bar{Q}^{(0)}_\alpha(\phi)\right)\delta'(\phi-\phi'),\cr
\tfrac{-1}{16\pi G}\left(\tfrac{2C_\rho}{d-1} \right)\left\{\bar{B}_a(\phi),B^{(0)}_c(\phi')\right\}&=&\bar{B}_b(\phi)f_{abc} + 2\delta_{ac}\delta'(\phi-\phi'),\cr
\tfrac{-1}{16\pi G}\left(\tfrac{2C_\rho}{d-1} \right)\left\{\bar{B}_a(\phi),\bar{Q}_{(0)}^\alpha(\phi')\right\}&=&-Q_\beta(\phi)(\Lambda_a)^{\beta\alpha}\delta(\phi-\phi'),\cr
\tfrac{-1}{16\pi G}\left(\tfrac{2C_\rho}{d-1} \right)\left\{B^{(0)}_a(\phi),B^{(0)}_c(\phi')\right\}&=&f_{abc}B^{(0)}_b(\phi)\delta(\phi-\phi'),\cr
\tfrac{-1}{16\pi G}\left(\tfrac{2C_\rho}{d-1} \right)\left\{B^{(0)}_a(\phi),Q_\alpha(\phi')\right\}&=&-Q_\beta(\phi)(\Lambda_a)^{\beta\alpha},\cr
\tfrac{-1}{16\pi G}\left(\tfrac{2C_\rho}{d-1} \right)\left\{B^{(0)}_a(\phi),\bar{Q}^{(0)}_\alpha(\phi')\right\}&=&-\bar{Q}^{(0)}_\beta(\phi)(\Lambda_a)^{\beta\alpha},\cr
\tfrac{-1}{16\pi G}\left(\tfrac{2C_\rho}{d-1} \right)\left\{B^{(0)}_a(\phi),\bar{B}_c(\phi')\right\}&=&f_{abc}\bar{B}_b(\phi)\delta(\phi-\phi')+2\delta_{ac}\delta(\phi-\phi'),\cr
\tfrac{i}{32\pi G}\left\{Q_\alpha(\phi),\bar{Q}^\beta(\phi')\right\}&=&\tfrac{1}{2}(\bar{B}_a(\phi)+\bar{B}_a(\phi'))(\Lambda^a)^\beta_{\,\,\alpha}\delta'(\phi-\phi')-\delta^\beta_\alpha\delta''(\phi-\phi')\cr
&&\hspace{2cm}\tfrac{1}{4}(2M(\phi)\delta^\beta_\alpha-\bar{B}_a(\phi)\bar{B}_b(\phi)(\Lambda^a)^\gamma_{\,\,\,\alpha}(\Lambda^b)^\beta_{\,\,\,\gamma})\delta(\phi-\phi'),\cr
\tfrac{i}{32\pi G}\left\{\bar{Q}^{(0)}_\alpha(\phi),\bar{Q}_{(0)}^\beta(\phi')\right\}&=&\tfrac{1}{2}(B^{(0)}_a(\phi)+B^{(0)}_a(\phi'))(\Lambda^a)^\beta_{\,\,\,\alpha}\delta'(\phi-\phi')\cr
&&\hspace{2cm}\tfrac{1}{2}(N^{(0)}(\phi)\delta^\beta_\alpha-B^{(0)}_a(\phi)\bar{B}_b(\phi)(\Lambda^a)^\gamma_{\,\,\,\alpha}(\Lambda^b)^\beta_{\,\,\,\gamma})\delta(\phi-\phi').\cr&&
\eea
Now we would like to shift 
\be
N^{(0)}\rightarrow N^{(0)}+\left(\tfrac{C_\rho}{d-1}\right)B_a\bar{B}^a\,\,\&\,\,M\rightarrow M+\left(\tfrac{C_\rho}{2(d-1)}\right)\bar{B}_a\bar{B}^a.
\ee
This is a shift by Sugawara tensor as it leaves invariant the Poisson brackets between $M$ and $N^{(0)}$, and between $N^{(0)}$ and $N^{(0)}$. The Poisson brackets then become
\bea
\tfrac{-1}{16\pi G}\left\{M(\phi),N^{(0)}(\phi')\right\}&=&-\delta'''(\phi-\phi')+(M(\phi)+M(\phi'))\delta'(\phi-\phi'),\cr
\tfrac{-1}{16\pi G}\left\{M(\phi),\bar{Q}^{(0)}_\alpha(\phi')\right\}&=&
 \left(\tfrac{1}{2}Q_\alpha(\phi')+Q_\alpha(\phi)\right)\delta'(\phi-\phi'),\cr
\tfrac{-1}{16\pi G}\left\{N^{(0)}(\phi),N^{(0)}(\phi')\right\}&=&
\left(N_{(0)}(\phi')+N_{(0)}(\phi)\right)\delta'(\phi-\phi'),\cr
\tfrac{-1}{16\pi G}\left\{N^{(0)}(\phi),Q_\alpha(\phi')\right\}&=&
\left(\tfrac{1}{2}Q_\alpha(\phi')+Q_\alpha(\phi)\right)\delta'(\phi-\phi'),\cr
\tfrac{-1}{16\pi G}\left\{N^{(0)}(\phi),\bar{Q}^{(0)}_\alpha(\phi')\right\}&=&
\left(\tfrac{1}{2}\bar{Q}^{(0)}_\alpha(\phi')+\bar{Q}^{(0)}_\alpha(\phi)\right)\delta'(\phi-\phi'),\cr
\tfrac{-1}{16\pi G}\left(\tfrac{2C_\rho}{d-1} \right)\left\{\bar{B}_a(\phi),B^{(0)}_c(\phi')\right\}&=&\bar{B}_b(\phi)f_{abc}\delta(\phi-\phi') + 2\delta_{ac}\delta'(\phi-\phi'),\cr
\tfrac{-1}{16\pi G}\left(\tfrac{2C_\rho}{d-1} \right)\left\{\bar{B}_a(\phi),\bar{Q}_{(0)}^\alpha(\phi')\right\}&=&-Q_\beta(\phi)(\Lambda_a)^{\beta\alpha}\delta(\phi-\phi'),\cr
\tfrac{-1}{16\pi G}\left(\tfrac{2C_\rho}{d-1} \right)\left\{B^{(0)}_a(\phi),B^{(0)}_c(\phi')\right\}&=&f_{abc}B^{(0)}_b(\phi)\delta(\phi-\phi'),\cr
\tfrac{-1}{16\pi G}\left(\tfrac{2C_\rho}{d-1} \right)\left\{B^{(0)}_a(\phi),Q_\alpha(\phi')\right\}&=&-Q_\beta(\phi)(\Lambda_a)^{\beta\alpha}\delta(\phi-\phi'),\cr
\tfrac{-1}{16\pi G}\left(\tfrac{2C_\rho}{d-1} \right)\left\{B^{(0)}_a(\phi),\bar{Q}^{(0)}_\alpha(\phi')\right\}&=&-\bar{Q}^{(0)}_\beta(\phi)(\Lambda_a)^{\beta\alpha}\delta(\phi-\phi'),\cr
\tfrac{-1}{16\pi G}\left(\tfrac{2C_\rho}{d-1} \right)\left\{B^{(0)}_a(\phi),\bar{B}_c(\phi')\right\}&=&f_{abc}\bar{B}_b(\phi)\delta(\phi-\phi')+2\delta_{ac}\delta(\phi-\phi'),\cr
\tfrac{i}{32\pi G}\left\{Q_\alpha(\phi),\bar{Q}^\beta(\phi')\right\}&=&\tfrac{1}{2}(\bar{B}_a(\phi)+\bar{B}_a(\phi'))(\Lambda^a)^\beta_{\,\,\alpha}\delta'(\phi-\phi')-\delta^\beta_\alpha\delta''(\phi-\phi')\cr
&&\hspace{0.2cm}\tfrac{1}{4}\left[2M(\phi)\delta^\beta_\alpha-\left(\tfrac{C_\rho}{d-1}\delta_\alpha^\beta+(\Lambda^{ab})_{\,\,\,\alpha}^\beta\right)\bar{B}_a(\phi)\bar{B}_b(\phi)\right]\delta(\phi-\phi'),\cr
\tfrac{i}{32\pi G}\left\{\bar{Q}^{(0)}_\alpha(\phi),\bar{Q}_{(0)}^\beta(\phi')\right\}&=&\tfrac{1}{2}(B^{(0)}_a(\phi)+B^{(0)}_a(\phi'))(\Lambda^a)^\beta_{\,\,\,\alpha}\delta'(\phi-\phi')\cr
&&\hspace{0.2cm}\tfrac{1}{2}\left[N^{(0)}(\phi)\delta^\beta_\alpha-\left(\tfrac{C_\rho}{d-1}\delta^\beta_\alpha+\tfrac{1}{2}\{\Lambda^a,\Lambda^b\}_{\,\,\,\alpha}^\beta\right)B^{(0)}_a(\phi)\bar{B}_b(\phi)\right]\delta(\phi-\phi').\cr&&
\eea 
The modes of which satisfy the following algebra in terms of Dirac brackets\footnote{Here we have absorbed the factors of $\left(\tfrac{1}{16\pi G}\right)$ and $\left(\tfrac{2C_\rho}{d-1}\right)$ into the charges and dropped the script$^{(0)}$. }
\bea
[M_m,N_n]&=&(m-n)M_{m+n}+\tfrac{1}{8G}m^3\delta_{m+n,0},
\hspace{0.7cm}[\bar{B}^a_m,B^b_n]=if^{abc}B^c_{m+n}+\tfrac{C_\rho}{2G(d-1)}m\delta^{ac}\delta_{m+n},\cr
[N_m,N_n]&=&(m-n)N_{m+n},
\hspace{3.4cm}[B^a_m,B^b_n]=if^{abc}B^c_{m+n},\cr
[M_m,\bar{Q}_{\alpha\, n}]&=&(\tfrac{m}{2}-n)Q_{\alpha\,(m+n)},
\hspace{2.95cm}[\bar{B}^a_m,\bar{Q}_{\alpha\,n}]=i(\Lambda^a)^{\alpha\beta}Q_{\beta\,(m+n)}\cr
[N_m,Q_{\alpha\,n}]&=&(\tfrac{m}{2}-n)Q_{\alpha\,(m+n)},
\hspace{2.95cm}[B^a_m,Q_{\alpha\,n}]=i(\Lambda^a)^{\alpha\beta}Q_{\beta\,(m+n)},\cr
[N_m,\bar{Q}_{\alpha\,n}]&=&(\tfrac{m}{2}-n)\bar{Q}_{\alpha\,(m+n)},
\hspace{2.95cm}[B^a_m,\bar{Q}_{\alpha\,n}]=i(\Lambda^a)^{\alpha\beta}\bar{Q}_{\beta\,(m+n)},\cr
&&\cr
\tfrac{1}{2}[Q_{\alpha\,m},\bar{Q}^{\beta}_n]&=&i(m-n)(\Lambda^a)^\beta_{\,\,\,\alpha}(\tfrac{d-1}{2C_\rho})\bar{B}^a_{m+n}+\tfrac{1}{8G}m^2\delta^\beta_\alpha\delta_{m+n,0}\cr
&&\hspace{0.2cm}+\tfrac{1}{4}\left[2M_{m+n}\delta^\beta_\alpha-\left(\tfrac{C_\rho}{d-1}\delta^\beta_\alpha+(\Lambda^{ab})^\beta_{\,\,\,\alpha}\right)16\pi G(\tfrac{d-1}{2C_\rho})^2(\bar{B}^a\bar{B}^b)_{m+n}\right],\cr
\tfrac{1}{2}[\bar{Q}_{\alpha\,m},\bar{Q}^{\beta}_n]&=&i(m-n)(\Lambda^a)^\beta_{\,\,\,\alpha}(\tfrac{d-1}{2C_\rho})B^a_{m+n}\cr
&&\hspace{0.2cm}+\tfrac{1}{2}\left[N_{m+n}\delta^\beta_\alpha-\left(\tfrac{C_\rho}{d-1}\delta^\beta_\alpha+\tfrac{1}{2}\{\Lambda^a,\Lambda^b\}^\beta_{\,\,\,\alpha}\right)16\pi G(\tfrac{d-1}{2C_\rho})^2(B^a\bar{B}^b)_{m+n}\right].\cr&&
\label{superbms3}
\eea
The nonlinearity on the the $r.h.s$ of the anti-commutator of fermionic charges is proportional to $\left(\tfrac{C_\rho}{d-1}\delta^\beta_\alpha+\tfrac{1}{2}\{\Lambda^a,\Lambda^b\}^\beta_{\,\,\,\alpha}\right)$ which is the same as the one that occurs in the super-conformal algebra obtained in \cite{Henneaux:1999ib}. In (\ref{superbms3})
$m,n$ in the fermionic charges $Q$ and $\bar{Q}$ being integers correspond to the Ramond sector, while $m,n$ being half-integers correspond to Neveau-Schwarz sector. The NS sector admits the full global sub-algebra spanned by $\{M_0,N_0,B^a_0,\bar{B}^a_0,Q^\alpha_{\pm\tfrac{1}{2}},\bar{Q}^\alpha_{\pm\tfrac{1}{2}}\}$, while the Ramond sector only contains sub-algebra spanned by $\{M_0,N_0,B^a_0,\bar{B}^a_0\}$.
\newline

\section{Conclusion}
In this paper we have surveyed, generalised and studied supersymmetric extensions of the boundary conditions of $AdS_3$ and ${\mathbb R}^{1,2}$ supergravities and their asymptotic symmetry algebras. 

In the first part of this paper we studied the generalisation of the non-dirichlet type boundary conditions introduced in \cite{Avery:2013dja, Troessaert:2013fma} to $AdS_3$ (extended) supergravity contexts. The extension of \cite{Avery:2013dja} reveals the existence of Ka\^{c}-Moody current of the relevant extended superalgebra concerned which is linear in its generators. This is not the case for the super-Virasoro algebra found in extended supergravity in \cite{Henneaux:1999ib}. In studying the chiral case we do uncover the Ward identities of the boundary chiral induced extended super-gravities which occur in \cite{Grisaru:1987pf, HariDass:1988dp}. The supersymmetrization of the boundary conditions studied here should be a first step in enabling one to see how such non-dirichlet boundary conditions occur in a string theoretic settings.  

In the second part we generalised asymptotically flat boundary conditions of ${\mathbb R}^{1,2}$ gravity to, again (extended) supergravity theories. In the latter case we found superalgebras containing $BMS_3$ that are nonlinear. These are the analogs of the nonlinear superalgebras of 2d CFTs studied long time ago \cite{Knizhnik:1986wc, Bershadsky:1986ms, Schoutens:1988tg, Defever:1991tf, Fradkin:1991gj, Fradkin:1992km, Fradkin:1992bz, Fradkin:1992cb, Bowcock:1992bm, Bina:1997bm, Ishimoto:1998hd}. The nonlinear extensions of Virasoro algebra found some applications in the AdS/CFT context before (see for instance \cite{Kraus:2007vu}). It is conceivable that the nonlinear BMS$_3$ superalgebras of the kind uncovered here will find suitable applications. A systematic classification of superalgebras containing $BMS_3$ is still under progress. The ${\mathbb R}^{1,2}$ supergravity calculation above predicts/suggests the existence of such nonlinear superalgebras whose large-$k$ limits are the ones found here. It will be interesting to uncover such nonlinear superalgebras on the lines of \cite{Knizhnik:1986wc, Bershadsky:1986ms, Schoutens:1988tg, Defever:1991tf, Fradkin:1991gj, Fradkin:1992km, Fradkin:1992bz, Fradkin:1992cb, Bowcock:1992bm, Bina:1997bm, Ishimoto:1998hd}. There are other types of boundary conditions for ${\mathbb R}^{1,2}$ gravity such as \cite{Grumiller:2017sjh, Fuentealba:2017omf} and it will be interesting to generalise such boundary conditions also to supergravity contexts.

In section \ref{liouville} in the Appendices we imposed conformal boundary conditions such that the conformal factor of the boundary metric obeys the Liouville equation $\partial_+\partial_-\log F=2\chi F$, which in general allows for non-vanishing boundary curvature. We uncovered the asymptotic symmetry algebra consisting of two copies of Virasoro corresponding to BH boundary conditions and two more copies of Virasoro with $c=-3\ell/3G$ corresponding to the stress-tensor modes of the Liouville field $F$ on the boundary. More recently people have uncovered many interesting generalisations of the non-dirichlet boundary conditions considered in the text. For example the ones in \cite{Perez:2016vqo} give the KdV equation as the bulk equation of motion. The Liouville case presented here is in similar spirit. It should be possible to extend these boundary conditions to supersymmetric contexts too. 
\section*{Acknowledgements}
We would like to thank Gautam Mandal for his feedback on some parts of the work presented here. RP would like to thank IMSc. for hospitality where part of this work was done.

\appendix

\section{Generalization to extended $AdS_3$ supergravity}
\label{Appendixextendedsugra}
In this appendix we give the detailed analysis of generalizing the chiral induced boundary conditions introduced in \cite{Avery:2013dja} to extended supergravity in $AdS_3$. Here, the left moving gauge field $\Gamma$ obeys the boundary conditions of the Dirichlet type studied in \cite{Henneaux:1999ib} and we repeat their analysis as it is for the left sector while imposing generalisation of chiral boundary condition on the right moving gauge field $\tilde{\Gamma}$. 
\vskip .5cm
\noindent\underline{\bf Conventions:}
\vskip .5cm
\label{Appendixextendedsugraconvention}
We follow the conventions of \cite{Henneaux:1999ib}. The structure constants for the $\tilde{G}$ are $f_{abc}$ which are completely anti-symmetric. The representation $\rho$ has the basis $(\lambda^a)^\alpha_{\,\,\,\beta}$ where $a$ counts the dimension of $\tilde{G}$ $i.e.$ $D$. Therefore, $[\lambda^a,\lambda^b]=f^{ab}_{\hspace{0.3cm} c}\lambda^c$. the Killing metric on $\tilde{G}$ is denoted by $g^{ab}=-f^{acd}f^{bcd}=-C_\nu\delta^{ab}$, where $C_\nu$ is the eigenvalue of the second Casimir in the adjoint representation of $\tilde{G}$. Similarly $tr(\lambda^a\lambda^b)=-\frac{d}{D}C_\rho\delta^{ab}$, where $C_\rho$ is the eigenvalue of the second Casimir in the representation $\rho$. We denote by $\eta^{\alpha\beta}$ the $\tilde{G}-$invariant symmetric metric on the representation $\rho$ which is orthogonal. Its inverse is $\eta_{\alpha\beta}$, this is used to raise and lower the supersymmetric (Greek) indices.

The list of all possible super-gravities in $AdS_3$ is given in \cite{Henneaux:1999ib}; we consider any such generic extended sugra in $AdS_3$.  
Below we list all the super-algebra generators:
\begin{itemize}
\item The $sl(2,{\mathbb R})$ generators are denoted as before by $(\sigma^0,\sigma^\pm)$
\bea
\sigma^0&=&\tfrac{1}{2}\sigma^3,\cr
\left[ \sigma^0,\sigma^{\pm} \right]&=&\pm \sigma^{\pm},\cr
\left[ \sigma^+,\sigma^- \right]&=&2\sigma^0.
\eea
The Killing form on $sl(2,{\mathbb R})$ is:
\be
Tr(\sigma^a\sigma^b)=h^{ab}=\frac{1}{4}
\begin{pmatrix}
2&0&0\\
0&0&4\\
0&4&0
\end{pmatrix}
\ee
\item The generators of $\tilde{G}$  which commute with $\sigma^a$s are\footnote{The indices on $\sigma$ always run over $(0,+,-)$ while those on $T$ run from $(1,\cdots,D)$, this is to be understood from the context.}:
\bea
\left[ T^a,T^b \right]&=&f^{ab}_{\hspace{0.3cm} c}\,\,T^c,\quad{\rm where}\,\,\,a\in\left\{ 1,D \right\},\cr
\left[ T^a,\sigma^b \right]&=&0,\qquad\quad\,\,\,{\rm where}\,\,\,b\in\left\{ +,-,0 \right\},\cr
STr\left(T^aT^b  \right)&=&\tfrac{2C_\rho}{d-1}\delta^{ab}.
\eea
\item The fermionic generators are denoted by $R^{\pm \alpha}$, where $\pm$ denotes the spinor indices with respect to the $sl(2,{\mathbb R})$ and $\alpha$ (Greek indices) denotes the vector index in the  representation $\rho$ of $\tilde{G}$.
\bea
\left[ \sigma^0,R^{\pm\alpha} \right]&=&\pm\tfrac{1}{2}R^{\pm \alpha},\quad{\rm where}\, \, \alpha\epsilon\left\{ 1,\cdots,d \right\},\cr
\left[ \sigma^{\pm},R^{\pm\alpha} \right]&=&0,\cr
\left[ \sigma^{\pm},R^{\mp\alpha} \right]&=&R^{\pm\alpha},\cr
\left[ T^a,R^{\pm\alpha} \right]&=&-(\lambda^a)^\alpha_{\,\,\beta}R^{\pm\beta},\cr
\left\{ R^{\pm\alpha},R^{\pm\beta} \right\}&=&\pm\eta^{\alpha\beta}\sigma^{\pm},\cr
\left\{ R^{\pm\alpha},R^{\mp\beta} \right\}&=&-\eta^{\alpha\beta}\sigma^0\pm\tfrac{d-1}{2C_\rho}(\lambda^a)^{\alpha\beta}T^a,\cr
&&\cr
STr\left( R^{-\alpha}R^{+\beta} \right)\!\!&=&\!\!-STr\left( R^{+\alpha}R^{-\beta} \right)=\eta^{\alpha\beta}.
\eea
Since the underlining algebra is now promoted to a graded Lie algebra, its generators satisfy the generalized Jacobi identity. The three-fermion Jacobi identity thus yields an identity for the matrices in the  representation $\rho$ of the internal algebra $\tilde{G}$:
\be
(\lambda^a)^{\alpha\beta}(\lambda^a)^{\gamma\delta}+(\lambda^a)^{\gamma\beta}(\lambda^a)^{\alpha\delta}=\frac{C_\rho}{d-1}(2\eta^{\alpha\gamma}\eta^{\beta\delta}-
\eta^{\alpha\beta}\eta^{\gamma\delta}-\eta^{\gamma\beta}\eta^{\alpha\delta})
\ee
\end{itemize}
The super-traces defined above are consistent, invariant and non-degenerate with respect to the super-algebra defined above and would be used in defining the action and the charges.

\subsection*{The Action}
\label{Appendixextendedsugraaction}
The super Chern-Simons action is defined as:
\be
S_{CS}[\Gamma]=\frac{k}{2\pi}\int_{\mathcal M}Str[\Gamma\wedge d\Gamma+\frac{2}{3}\Gamma\wedge\Gamma\wedge\Gamma].
\ee 
The above integration is over a three manifold ${\mathcal M}=D\times{\mathbb R}$, where $D$ has a topology of a disk. The level $k$ of the Chern-Simons action is related to the Newton's constant $G$ in three dimension and the $AdS$ length $\ell$ through $k=\ell/(4G)$. The product of two fermions differs by a factor of $i$ from the standard Grasmann product ($(ab)^*=b^*a^*$).This basically requires one the multiply a factor of $-i$ where ever $\eta^{\alpha\beta}$ occurs, and where ever $\tfrac{d-1}{2C\rho}(\lambda^a)^{\alpha\beta}$ occurs while evaluating anti-commutator between fermionic generators  in the calculations below\footnote{This basically so because the product of two real Grasmann fields is imaginary. This is equivalent to using
\bea
\left\{ R^{\pm\alpha},R^{\pm\beta} \right\}&=&\mp i\eta^{\alpha\beta}\sigma^{\pm},\cr
\left\{ R^{\pm\alpha},R^{\mp\beta} \right\}&=&i\eta^{\alpha\beta}\sigma^0\mp i\tfrac{d-1}{2C_\rho}(\lambda^a)^{\alpha\beta}T^a,\cr
&&\cr
STr\left( R^{-\alpha}R^{+\beta} \right)\!\!&=&\!\!-STr\left( R^{+\alpha}R^{-\beta} \right)=-i\eta^{\alpha\beta},
\eea
instead of the one stated in the commutation relations of the extended super-algebra.
}.
In the Chern-Simons formulation of (super-)gravity, the metric (and other fields) which occur in Einstein-Hilbert (Hilbert-Palatini) action are a derived concept. The equations of motion for the Chern-Simons action can for example be satisfied by gauge field configurations which may yield a non-singular metric. There fore one has to make sure that such configurations are not considered in the analysis.

The supergravity action for the above super-algebra can be written in full detail yielding the action in the Hilbert-Palatini form:
\bea
S[\Gamma,\tilde{\Gamma}]&=&\tfrac{1}{8\pi G}\int_{\mathcal M}d^3x \{\tfrac{1}{2}eR+\frac{e}{\ell^2}+\cr
&&-\frac{i\ell}{2}\varepsilon^{ijk}(\psi_i){\mathcal D}^{\mu\nu}_j(\psi_k)_\nu+\frac{i\ell}{2}\varepsilon^{ijk}(\tilde{\psi}_i)\tilde{{\mathcal D}}^{\mu\nu}_j(\tilde{\psi}_k)_\nu \}\cr
&&+\frac{C_\rho}{d-1}\ell\varepsilon^{ijk}(B^a_i\partial_j B^a_k +\tfrac{1}{3}f_{abc}B^a_iB^b_jB^c_k)\cr
&&-\frac{C_\rho}{d-1}\ell\varepsilon^{ijk}(\tilde{B}^a_i\partial_j \tilde{B}^a_k +\tfrac{1}{3}f_{abc}\tilde{B}^a_i\tilde{B}^b_j\tilde{B}^c_k)\cr
&&-\frac{i}{2}\varepsilon^{ijk}\eta^{\alpha\beta}e^a_i([\bar{\psi}_j]_\alpha t^a[\psi_k]_\beta-[\tilde{\bar{\psi}}_j]_\alpha t^a[\tilde{\psi}_k]_\beta)\}
\eea  
The square brackets denote the two-component $sl(2,{\mathbb R})$ spinor representations. The spin covariant operators ${\mathcal D}$ and $\tilde{\mathcal D}$ are:
\bea
{\mathcal D}^{\mu\nu}_j&=&
\begin{pmatrix}
2\,\big(\, \eta^{\alpha\beta}\partial_j+(\lambda^a)^{\alpha\beta}B^a_j\,\big)\,\,
\delta^\mu_{+\alpha}\delta^\nu_{-\beta}+\\
-\eta^{\alpha\beta}\big(\,\tfrac{1}{2}\omega^3_j[\delta^\mu_{+\alpha}\delta^\nu_{-\beta}+
\delta^\mu_{-\alpha}\delta^\nu_{+\beta}]+
\omega^+_j\delta^\mu_{-\alpha}\delta^\nu_{-\beta}
-\omega^-_j\delta^\mu_{+\alpha}\delta^\nu_{+\beta} \, \big     )
\end{pmatrix},\cr&&\cr&&\cr
\tilde{{\mathcal D}}^{\mu\nu}_j&=&
\begin{pmatrix}
2\,\big(\, \eta^{\alpha\beta}\partial_j+(\lambda^a)^{\alpha\beta}\tilde{B}^a_j\,\big)\,\,
\delta^\mu_{+\alpha}\delta^\nu_{-\beta}+\\
-\eta^{\alpha\beta}\big(\,\tfrac{1}{2}\omega^3_j[\delta^\mu_{+\alpha}\delta^\nu_{-\beta}+
\delta^\mu_{-\alpha}\delta^\nu_{+\beta}]+
\omega^+_j\delta^\mu_{-\alpha}\delta^\nu_{-\beta}
-\omega^-_j\delta^\mu_{+\alpha}\delta^\nu_{+\beta} \, \big     )
\end{pmatrix}
\eea
From the form of the above action it is quite evident that the analysis done in the Hilbert-Palatini formulation of supergravity would be quite cumbersome if not difficult. Further, it was found that computation of the asymptotic charge associated with gauge transformations which vary the super-gauge field at the $AdS$ asymptote\footnote{By this we mean the boundary of the disk $D$} via the prescription of Barnich $et\,\,al$ \cite{Barnich:2001jy} for the above form of the action is too difficult. The same prescription of computing asymptotic charges in the Chern-Simons formalism yields a know expression for asymptotic charge in Chern-Simons theory. Therefore we proceed as before with the analysis in the Chern-Simons prescription.   

\subsection{Boundary conditions}
\label{Appendixextendedsugraboundary}

The fall-off conditions in terms of the gauge fields are:
\bea
\Gamma &=& bdb^{-1}+bab^{-1},\cr
\tilde{\Gamma}&=& b^{-1}db + b^{-1}\tilde{a}b,\cr
{\rm where}\,\,\,b&=&e^{\sigma^0\ln(r/\ell)},\cr
a&=&\left[  \sigma^- + L\sigma^+ +\psi_{+\alpha +}R^{+\alpha} + B_{a+}T^a    \right]dx^+,\cr
\tilde{a}&=&\left[\sigma^+ + \bar{L}\sigma^-+\bar{\psi}_{-\alpha-}R^{-\alpha}+\bar{B}_{a-} T^a \right]dx^-\cr&&+\left[ \tilde{A}_{a+}\sigma^a + \tilde{B}_{a+}T^a + \tilde{\psi}_{+\alpha+}R^{+\alpha}+ \tilde{\psi}_{-\alpha+}R^{-\alpha} \right]dx^+.
\eea
Here the $dx^-$ component of the gauge field $\tilde{a}$ one form is that of a super-gauge field corresponding to Dirichlet boundary condition as given in \cite{Henneaux:1999ib}. All functions above are ${\it a\, priori}$ functions of both the boundary coordinates. The equation of motion- as mentioned earlier, is implied by the flatness condition imposed on the two gauge fields. For the right gauge field this implies that the functions are independent of the $x^-$ co-ordinate. $i.e.$ $\partial_-a=0$. 
\be
\partial_-L=\partial_-\psi_{+\alpha +}=\partial_-B_{a+}=0
\ee
For the left gauge field we would like to use the equations of motion to solve for the $\tilde{a}_+$ components. This gives the $\tilde{a}_+$ components in terms of $\tilde{A}_{++},\tilde{B}_{a+}$, $\tilde{\psi}_{+\alpha +}$ and the $\tilde{a}_-$ components: 
\bea
\tilde{A}_{0+}&=&\partial_-\tilde{A}_{++},\cr
\tilde{A}_{-+}&=&\tilde{A}_{++}\bar{L} - \tfrac{1}{2}\partial^2_-\tilde{A}_{++}+i\tfrac{\eta^{\alpha\beta}}{2}\tilde{\psi}_{+\alpha+}\bar{\psi}_{-\beta-},\cr
\tilde{\psi}_{-\alpha+}&=&\tilde{A}_{++}\bar{\psi}_{-\alpha-}-\partial_-\tilde{\psi}_{+\alpha+} + (\lambda^a)^\beta_{\,\,\alpha}\bar{B}_a\tilde{\psi}_{+\beta+}.
\eea 
Provided they satisfy the following set of differential equations:
\bea
\partial_+\bar{L}+\tfrac{1}{2}\partial^3_-\tilde{A}_{++}-2\bar{L}\partial_-\tilde{A}_{++}-
\tilde{A}_{++}\partial_-\bar{L}
&&\cr+i\eta^{\alpha\beta}\bar{\psi}_{-\beta-}\left( \tilde{A}_{++}\bar{\psi}_{-\alpha-} + (\lambda^a)^\beta_{\,\,\alpha}\bar{B}_{a-}\tilde{\psi}_{+\beta+}  + \partial_-\tilde{\psi}_{+\alpha+} \right) +i \eta^{\alpha\beta}\partial_-(\bar{\psi}_{-\beta-}\tilde{\psi}_{+\alpha+})&=&0,\cr
\partial_+\bar{B}_{a-}-\partial_-\tilde{B}_{a+}+f^{bc}_{\,\,\,\,\,a}\tilde{B}_{b+}\bar{B}_{c-}+i\tfrac{d-1}{2C_\rho}(\lambda^a)^{\alpha\beta}\tilde{\psi}_{+\alpha+}\bar{\psi}_{-\beta-}&=&0,\cr
\partial_+\bar{\psi}_{-\alpha-} -\partial_-[\tilde{A}_{++}\bar{\psi}_{-\alpha-} - \partial_-\tilde{\psi}_{+\alpha+} + (\lambda^a)^\beta_{\,\,\alpha}\bar{B}_{a-}\tilde{\psi}_{+\beta+}] - \tfrac{1}{2}\partial_-\tilde{A}_{++}\bar{\psi}_{-\alpha-} &&\cr+ (\lambda^a)^\beta_{\,\,\alpha}\bar{B}_{a-}[\tilde{A}_{++}\bar{\psi}_{-\beta-} - \partial_-\tilde{\psi}_{+\beta+} + (\lambda^a)^\gamma_{\,\,\beta}\bar{B}_{b-}\tilde{\psi}_{+\gamma+}] -(\lambda^a)^\beta_\alpha\tilde{B}_{a+}\bar{\psi}_{-\beta-}
-\bar{L}\tilde{\psi}_{+\alpha+}&=&0.\nonumber\\
\eea
These are the Ward identities expected to be satisfied by the induced gravity theory on the boundary. 
We will later choose the $\bar{()}$ functions such that global $AdS_3$ is a part of the moduli space of bulk solutions.$i.e.$ $\bar{L}=\tfrac{-1}{4}$ and $\bar{B}=0=\bar{\psi}$.\\\\ 
In the following analysis we will consider the sources $i.e.$ the bared functions as constants along the boundary directions. There is no need to assume this, and we have done so only for simplicity in the expressions for change in the moduli space parameters. Either ways, demanding that the bared functions- $\bar{L},\bar{B},\bar{\psi}$, be treated as sources which determine aspects of the theory requires adding of specific boundary term to the bulk action. As explained previously, this is done so that the required set of bulk solutions obey the variational principle.\\\\
The boundary term to be added is given by:
\bea
S_{bndy}&=&\frac{k}{8\pi}\int_{\mathcal \partial M}d^2x\,\, STr(-\sigma^0[\tilde{a}_+,\tilde{a}_-])-2\bar{L}_0\sigma^-\tilde{a}_+ + (\tfrac{d-1}{2C_\rho})^2T^aT^bSTr(\tilde{a}_+T_a)STr(\tilde{a}_-T_b)\cr
&&\hspace{1.5cm} -2(\tfrac{d-1}{2C_\rho})\bar{B}_{0a}T^aT^bSTr(\tilde{a}_+T^b)-\tfrac{1}{2}(\bar{\psi}_0)_{-\alpha}R^{-\alpha}\tilde{a}_+ ).
\eea
This implies the following desired variation of the total action:
\bea
\delta S_{total}&=&\frac{k}{8\pi}\int_{\mathcal M}\!\!d^2x\,\,2(\bar{L}-\bar{L}_0)\delta\tilde{A}_{++}+2(\tfrac{2C_\rho}{d-1})(\bar{B}_{a-}-\bar{B}_{0a})\delta\tilde{B}_{a+}+\tfrac{i}{2}(\bar{\psi}_{-\alpha-}
-(\bar{\psi}_0)_{-\alpha})\delta\tilde{\psi}_{+\alpha+}\eta^{\alpha\beta}
\nonumber\\
\eea
 In our present case, we would be choosing the later by fixing $\bar{L}=-1/4,\bar{B}_{0a}=0=(\bar{\psi}_0)_{-\alpha}$. Thus the variational principle is satisfied for configurations with $\bar{L}=\tfrac{-1}{4}$ and $\bar{B}_{a-}=0=\bar{\psi}_{-\alpha-}$ which describes global $AdS_3$.



\subsection{Charges and symmetries}
Just as in the previous sections, one needs to find the space of gauge transformations that maintains the above form of the gauge fields, thus inducing transformations on the functions $\tilde{A}_{a+},\tilde{B}_{a+},\tilde{\psi}_{+\alpha+},L,B_a,\psi_{+\alpha+}$ which parametrize the space of solutions. Once this is achieved, one can define asymptotic conserved charge associated with the change induced by such residual gauge transformations on the space of solutions. For the boundary conditions to be well defined, this asymptotic charge must be finite and be integrable on the space of solutions.
\subsubsection*{Right sector}
\label{Appendixextendedsugraright}
The analysis of the left sector $i.e.$ on the gauge field $\Gamma$ is exactly the one done in \cite{Henneaux:1999ib}. Here we analyze the right sector. For the choice of $\bar{L}=-\tfrac{1}{4},\bar{B}=0=\bar{\psi}$ the $eom$ can be solved and the solutions can be  parametrized as below:
\bea
\tilde{A}_{++}&=&f(x^+)+g(x^+)e^{ix^-}+\bar{g}(x^+)e^{-ix^-},\cr
\tilde{B}_{a+}&\cong&\tilde{B}_{a+}(x^+),\cr
\tilde{\psi}_{+\alpha+}&=&\chi_\alpha(x^+)e^{ix^-/2}+\bar{\chi}_\alpha(x^+)e^{-ix^-/2}.
\eea
We would now seek the residual gauge tranformation parameters that would keep the above form of the gauge field $\tilde{\Gamma}$ form invariant. The residual gauge transformations are generated by $\tilde{\Lambda}=\xi_a\sigma^a+b_aT^a+\epsilon_{+\alpha}R^{+\alpha}
+\epsilon_{-\alpha}R^{-\alpha}$ with the constraint that $\delta\tilde{a}_-=0$:
\bea
\delta\tilde{a}_-&=&d\tilde{\Lambda} +[\tilde{a}_-,\tilde{\Lambda}],\cr
\implies\xi_0&=&\partial_-\xi_+,\cr
\xi_-&=&-\tfrac{1}{4}(1+2\partial_-^2)\xi_+,\cr
\epsilon_{-\alpha}&=&-\partial_-\epsilon_{+\alpha},\cr
\partial_-(1+\partial^2_-)\xi_+&=&0,\cr
\partial_-b_a=&0&=(\partial^2_-+\tfrac{1}{4})\epsilon_{+\alpha}.
\eea
One can solve for the residual gauge transformations:
\bea
\xi_+&=&\lambda_f(x^+)+\lambda_g(x^+)e^{ix^-}+\bar{\lambda}_{\bar{g}}(x^+)e^{-ix^-},\cr
b_a&\cong&b_a(x^+),\cr
\epsilon_{+\alpha}&=&\varepsilon_\alpha(x^+)e^{ix^-/2}+\bar{\varepsilon}_\alpha(x^+)e^{-ix^-/2}.
\eea 
Here too, one finds that the functions parametrizing the space of solutions and residual gauge transformations are functions of $x^+$ alone. The $x^+$ dependence of the functions will be suppressed from here on for neatness. The variation of the above parameters under the residual gauge transformations are:
\bea
\delta f&=&\lambda_f' + 2i(g\bar{\lambda}_{\bar{g}}-\bar{g}\lambda_g)+i\eta^{\alpha\beta}(\chi_\alpha\bar{\varepsilon}_\beta+\bar{\chi}_\alpha\varepsilon_\beta),\cr
\delta g&=&\lambda_g' +i(g\lambda_f-\lambda_g f)+i\eta^{\alpha\beta}\chi_\alpha\varepsilon_\beta,\cr
\delta\bar{g}&=&\bar{\lambda}_{\bar{g}}' -i(\bar{g}\lambda_f-\bar{\lambda}_{\bar{g}}f)+i\eta^{\alpha\beta}\bar{\chi}_\alpha\bar{\varepsilon}_\beta,\cr
\delta\tilde{B}_{a+}&=&b_a' + f_a^{\,\,bc}\tilde{B}_{b+}b_c + \tfrac{d-1}{2C_\rho}(\lambda_a)^{\alpha\beta}(\bar{\chi}_\alpha\varepsilon_\beta-\chi_\alpha\bar{\varepsilon}_\beta),\cr
\delta\chi_\alpha&=&\varepsilon_\alpha' - (\lambda^a)^\beta_{\,\,\alpha}[\tilde{B}_{a+}\varepsilon_\beta-b_a\chi_\beta] + i[g\bar{\varepsilon}_\alpha - \tfrac{f}{2}\varepsilon_\alpha - \lambda_g\bar{\chi}_\alpha +\tfrac{\lambda_f}{2}\chi_\alpha],\cr
\delta\bar{\chi}_\alpha&=&\bar{\varepsilon}_\alpha' - (\lambda^a)^\beta_{\,\,\alpha}[\tilde{B}_{a+}\bar{\varepsilon}_\beta-b_a\bar{\chi}_\beta] - i[\bar{g}\varepsilon_\alpha - \tfrac{f}{2}\bar{\varepsilon}_\alpha - \bar{\lambda}_{\bar{g}}\chi_\alpha +\tfrac{\lambda_f}{2}\bar{\chi}_\alpha]
\eea
The charges corresponding to these transformation is given by:
\bea
\mathrlap{\slash}\delta Q [\tilde{\Lambda}]&=& -\frac{k}{2\pi}\int d\phi\, Str[\tilde{\Lambda},\delta\tilde{a}_\phi].
\eea
The above charge can be integrated to
\bea
Q[\Lambda]&=&-\tfrac{k}{2\pi}\int d\phi [-\tfrac{f}{2}\lambda_f + g\bar{\lambda}_{\bar{g}} + \bar{g}\lambda_g +\tfrac{2C_\rho}{d-1}\tilde{B}_{a+}b^a + \eta^{\alpha\beta}(\chi_\alpha\bar{\varepsilon}_\beta-\bar{\chi}_\alpha\varepsilon_\beta)  ].
\eea
This charge is the generator of canonical transformations on the space of solutions parametrized by set of functions $F$ $via$ the Poisson bracket.
\bea
\delta_{\tilde{\Lambda}} F &=& \{Q[\tilde{\Lambda}],F \}
\eea
Therefore the Poisson bracket algebra is:
\bea
&&\{f({x^+}'),f(x^+)\}=-2\alpha_Q\delta'({x^+}'-x^+),\hspace{2cm}\{\chi_\alpha ({x^+}'),f(x^+) \}=-i\alpha_Q \delta({x^+}'-x^+)\chi_\alpha,\cr
&&\{g({x^+}'),f(x^+)\}=-2i\alpha_Q g(x^+)\delta({x^+}'-x^+),\hspace{1cm}\{\bar{\chi}_\alpha ({x^+}'),f(x^+) \}=i\alpha_Q \delta({x^+}'-x^+)\bar{\chi}_\alpha,\cr
&&\{\bar{g}({x^+}'),f(x^+)\}=2i\alpha_Q \bar{g}({x^+})\delta({x^+}'-x^+),\hspace{1.35cm}\{\bar{\chi}_\alpha ({x^+}'),g(x^+) \}=i\alpha_Q \delta({x^+}'-x^+)\chi_\alpha,\cr
&&\{\bar{g}({x^+}'),g(x^+)\}=i\alpha_Q f(x^+)\delta({x^+}'-x^+)+\alpha_Q \delta'({x^+}'-x^+),\cr
&&\{\chi_\alpha ({x^+}'),\bar{g}(x^+) \}=-i\alpha_Q \delta({x^+}'-x^+)\bar{\chi}_\alpha,\cr
&&\{\tilde{B}_{a+}({x^+}'),\tilde{B}_{b+}(x^+)\}=-\alpha_Q(\tfrac{d-1}{2C_\rho})\delta({x^+}'-x^+)f_{ab}^{\,\,\,\,c}\tilde{B}_{c+}(x^+)+\alpha_Q(\tfrac{d-1}{2C_\rho}) \delta'({x^+}'-x^+)\delta_{ab},
\eea
while those among the fermions is:
\bea
&&\{\bar{\chi}_\alpha({x^+}'),\chi_\beta(x^+)\}=\tfrac{i\alpha_Q}{2}\eta_{\alpha\beta}f(x^+)\delta({x^+}'-x^+)+\alpha_Q(\lambda^a)_{\alpha\beta}\tilde{B}_{a+}\delta({x^+}'-x^+)\cr
&&\hspace{4cm}+\alpha_Q\eta_{\alpha\beta}\delta'({x^+}'-x^+),\cr
&&\{\chi_\alpha({x^+}'),\chi_\beta(x^+)\}=i\alpha_Q \eta_{\alpha\beta}g(x^+)\delta({x^+}'-x^+),\cr
&&\{\bar{\chi}_\alpha({x^+}'),\bar{\chi}_\beta(x^+)\}=i\alpha_Q \eta_{\alpha\beta}\bar{g}(x^+)\delta({x^+}'-x^+),\cr
&&\{\tilde{B}_{a+}({x^+}'),\chi_\beta(x^+)\}=-\alpha_Q
(\tfrac{d-1}{2 C_\rho})(\lambda_a)^\alpha_{\,\,\beta}\chi_\alpha(x^+)\delta({x^+}'-x^+),\cr
&&\{\tilde{B}_{a+}({x^+}'),\bar{\chi}_\beta(x^+)\}=-\alpha_Q
(\tfrac{d-1}{2 C_\rho})(\lambda_a)^\alpha_{\,\,\beta}\bar{\chi}_\alpha(x^+)\delta({x^+}'-x^+).
\eea
where $\alpha_Q =\tfrac{2\pi}{k}$. Rescaling the above currents to:
\bea
&&f\rightarrow\tfrac{k}{4\pi}f,\hspace{1.1cm}g\rightarrow\tfrac{k}{2\pi}g,\hspace{0.8cm}\bar{g}\rightarrow\tfrac{k}{2\pi}\bar{g},\cr
&&\tilde{B}_{a+}\rightarrow\tfrac{k}{2\pi}\tilde{B}_{a+},\hspace{0.3cm}\chi_{\alpha}\rightarrow\tfrac{k}{2\pi}\chi_{\alpha},\hspace{0.4cm}\bar{\chi}_{\alpha}\rightarrow\tfrac{k}{2\pi}\bar{\chi}_{\alpha},
\eea
and expanding it in the modes yields the following commutators:
\bea
&&[f_m,f_n]=m\tfrac{k}{2}\delta_{m+n,0},\hspace{3.65cm}[(\chi_\alpha)_m,f_n]=\tfrac{1}{2}(\chi_\alpha)_{(m+n)},\cr
&&[g_m,f_n]=g_{m+n},\hspace{4.5cm}[(\bar{\chi}_\alpha)_m,f_n]=-\tfrac{1}{2}(\bar{\chi}_\alpha)_{(m+n)},\cr
&&[\bar{g}_m,f_n]=-\bar{g}_{m+n},\hspace{4.156cm}[(\bar{\chi}_\alpha)_m,g_n]=-(\chi_\alpha)_{m+n},\cr
&&[\bar{g}_m,g_n]=-2f_{m+n}-mk\delta_{m+n,0},\hspace{1.75cm}[(\chi_\alpha)_m,\bar{g}_n]=(\bar{\chi}_\alpha)_{m+n},\cr&&\cr
&&\{(\chi_\alpha)_m,(\chi_\beta)_n\}=-\eta_{\alpha\beta}g_{m+n},\hspace{2.2cm}\{(\bar{\chi_\alpha})_m,(\bar{\chi_\beta})_n\}=-\eta_{\alpha\beta}\bar{g}_{m+n},\cr
&&[(\tilde{B}_{a+})_m,(\chi_\beta)_n]=i(\tfrac{d-1}{2C_\rho})(\lambda_a)^\alpha_{\,\,\,\beta}(\chi_\alpha)_{(m+n)},\hspace{0.1cm}[(\tilde{B}_{a+})_m,(\bar{\chi}_\beta)_n]=i(\tfrac{d-1}{2C_\rho})(\lambda_a)^\alpha_{\,\,\,\beta}(\bar{\chi}_\alpha)_{(m+n)},\cr&&\cr
&&[(\tilde{B}_{a+})_m,(\tilde{B}_{b+})_n]=-i(\tfrac{d-1}{2C_\rho})f_{ab}^{\,\,\,\,c}(\tilde{B}_{c+})_{(m+n)}-(\tfrac{d-1}{2C_\rho})km\delta_{ab}\delta_{m+n,0},\cr
&&\{(\bar{\chi}_\alpha)_m,(\chi_\beta)_n\}=-\eta_{\alpha\beta}f_{(m+n)}+i(\lambda^a)_{\alpha\beta}(\tilde{B}_{c+})_{(m+n)}-km\eta_{\alpha\beta}\delta_{m+n,0}.
\eea
This is the affine Ka\v{c}-Moody super-algebra. 
Here, it is evident that the central extension to the $sl(2,{\mathbb R})$current sub-algebra spanned by ($f,g,\bar{g}$) is $k=c/6$. The quadratic nonlinearities that occur in the super-Virasoro are not present here. 

\section{Holographic Liouville theory}
\label{liouville}
In this appendix we begin with generalising the boundary conditions of \cite{Troessaert:2013fma} so as to let the boundary metric have a  non-vanishing curvature. The motivation is to provide a holographic description of the Liouville equation (instead of the free-field equation as in \cite{Troessaert:2013fma}). The boundary conditions in \cite{Troessaert:2013fma} for the metric components look like:


\begin{equation}
\begin{aligned}
g_{rr} &= \frac{l^2}{r^2} + {\cal O}(r^{-4}), ~~ g_{r+} = {\cal O}(r^{-1}), ~~ g_{r-} = {\cal O}(r^{-3}), \\
g_{+-} &= - \frac{r^2}{2}F(x^+, x^-) + {\cal O}(r^0), ~~ g_{--} =  {\cal O}(r^0), \\
g_{++} &=   {\cal O} (r^0) ,
\end{aligned}
\label{Liouville_bndy_cond}
\end{equation}
where $F(x^+,x^-)$ satisfies $\partial_+\partial_-\log F=0$, yielding the boundary metric to have zero curvature. In contrast we impose on $F$ the generic Liouville equation: $\partial_-\partial_+\log F=2\chi F$. 
Here $x^+, x^-$ are treated to be the boundary coordinates and $r$ is the radial coordinate with the asymptotic boundary at $r^{-1} = 0$. One can write a general nonlinear solution of $AdS_3$ gravity in Fefferman--Graham
coordinates  \cite{Skenderis:1999nb} as:
\begin{equation}
\label{nlsoln1}
ds^2 = l^2 \frac{dr^2}{r^2} + r^2 \left[ g^{(0)}_{ab} + \frac{l^2}{r^2} \, g^{(2)}_{ab} + \frac{l^4}{r^4} g^{(4)}_{ab} \right] dx^a dx^b.
\end{equation}
Therefore, the full set of nonlinear solutions consistent with our boundary conditions is obtained when
\begin{equation}
\label{nlsoln2}
\begin{aligned}
g^{(0)} _{++ } &= 0, ~~ g^{(0)}_{+-} = -\frac{1}{2} F(x^+, x^-), ~~ g^{(0)}_{--} = 0, \\
g^{(2)}_{++} &= \kappa(x^+, x^-), ~~  g^{(2)}_{+-} = \sigma(x^+,x^-), 
      ~~ g^{(2)}_{--} =\tilde \kappa (x^+, x^-),\\
g^{(4)}_{ab} &= \frac{1}{4} g^{(2)}_{ac} g_{(0)}^{cd} g^{(2)}_{db} \, ,
\end{aligned}\end{equation}
where in the last line $g_{(0)}^{cd}$ is $g^{(0)}_{cd}$ inverse. Imposing the equations of motion $R_{\mu\nu} - \frac{1}{2} R \, g_{\mu\nu} - \frac{1}{l^2} g_{\mu\nu}=0$ one finds that these equations are satisfied for $\mu, \nu = +, -$. Then the remaining three equations coming from $(\mu, \nu)  = (r,r), (r,+), (r,-)$ impose the following relations:
\bea
\label{nlsoln3}
\sigma (x^+, x^-) - \frac{1}{2} \partial_+ \partial_- \log F =0, ~~ \cr
\partial_- \kappa = F \, \partial_+ \left(\frac{\sigma}{F}\right), ~~ \partial_+ \tilde \kappa = F \, \partial_- \left(\frac{\sigma}{F} \right)
\eea
 In general the equations can be solved for $\kappa$ and $\tilde \kappa$ in terms of $F$ as follows:
\bea
\kappa (x^+, x^-) &=& \kappa_0 (x^+) + \frac{1}{2} \partial_+^2 \log F - \frac{1}{4} (\partial_+ \log F)^2 \cr
\tilde \kappa (x^+, x^-) &=& \tilde \kappa_0 (x^-) + \frac{1}{2} \partial_-^2 \log F - \frac{1}{4} (\partial_- \log F)^2
\eea
We now have to specialise to some subset of solutions such that we have Liouville equation satisfied by $F$. For this observe that when $\partial_- \kappa = \partial_+ \tilde \kappa = 0$ we have $\sigma = \chi \, F$ for some constant $\chi$. Then the ward identity $\sigma = \frac{1}{2} \partial_+\partial_- \log F$ reads:
\bea
\frac{1}{2} \partial_+\partial_- \log F = \chi \, F
\eea
which is the famous Liouville's equation. So if we add boundary terms such that we keep $\sigma = \chi \, F$ then it follows that $\partial_- \kappa = \partial_+ \tilde \kappa = 0$. For this, it is useful to note that the boundary (holographic) stress tensor $T_{ij}$ for the class of metrics we have is proportional to
\bea
g^{(2)}_{\mu\nu} - R^{(0)} \, g^{(0)}_{\mu\nu} = \left(\begin{array}{cc} \kappa & 5 \, \sigma \\ 5 \, \sigma & \tilde \kappa \end{array} \right)
\eea
Taking the trace with respect to the boundary metric gives
\bea
g_{(0)}^{\mu\nu} (g^{(2)}_{\mu\nu} - R^{(0)} \, g^{(0)}_{\mu\nu} ) = - 20 \frac{\sigma}{F}
\eea
Therefore the constraint $\sigma = \chi \, F$ simply translates into demanding $g_{(0)}^{\mu\nu} (g^{(2)}_{\mu\nu} - R^{(0)} \, g^{(0)}_{\mu\nu} ) = -20 \, \chi$.
The variation of the action along the solution space is
\bea
\delta S = \frac{1}{2} \int_{bdy.} d^2x \, \sqrt{|g^{(0)}|} ~T^{ij} \delta g^{(0)}_{ij} = -\frac{l}{8\pi \, G} \int_{bdy} d^2x \, \frac{5 \sigma}{F} \delta F.
\eea
So we add the boundary term:
\bea
\frac{l}{8\pi G} \int_{bdy.} d^2x \, 10 \chi \, \sqrt{|g^{(0)}|} = \frac{l}{8\pi \, G} \int_{bdy} d^2x \, 5 \chi \, F
\eea
such that the total variation of the action is
\bea
\delta S_{total} = -\frac{l}{8\pi \, G} \int_{bdy} d^2x \, 5 \, (\frac{\sigma}{F} - \chi) \, \delta F.
\eea
Now we could choose either $\delta F = 0$ (Dirichlet) or $\sigma = \chi \, F$ (Neumann). Choosing the latter gives rise to the Liouville equation as we desire.
\subsection{Classical Solutions and asymptotic symmetries}
It is well known that the general solution of the Liouville equation $\partial_+ \partial_- \log F = 2\chi \, F$ is given by
\bea
\label{Liouvillesol}
F = \chi^{-1} \,  \frac{\partial_+ f(x^+) \, \partial_- \tilde f (x^-)}{[1+f(x^+)  \, \tilde f (x^-)]^2} ~~ {\rm for } ~~ \chi \ne 0,  ~~~
F = f(x^+) \, \tilde f(x^-) ~~{\rm for } ~~ \chi = 0.
\eea
The $\chi=0$ case was considered by \cite{Troessaert:2013fma}. 
We now proceed to obtain the asymptotic symmetries for the above boundary conditions. The residual diffeomorphisms that the leave the metric in the above form are:
\bea
&&\xi = r\xi^r \partial_r + (\xi^++(\mathcal O)(\tfrac{1}{r})) \partial_+ + (\xi^-+{\mathcal O}(\tfrac{1}{r}))\partial_- ,\cr&&\cr
{\rm where} && \partial_-\xi^+ =\partial_+\xi^-=0 \,\,,\,\,\partial_+ \partial_- \xi^r = 2\xi^r F,
\label{residual_diff_Liouville}
\eea
the subleading functions in $r$ are all determined from the boundary values of components. For convenience of calculation, let us introduce a field $\Phi=\log (\chi F)$. The equation for $\Phi$ then is:
\be
\partial_+\partial_-\Phi=2e^{\Phi}.
\ee
The first order variation of the above differential equation satisfies:
\be
\partial_+\partial_-\delta\Phi=2e^\Phi \delta\Phi,
\ee
therefore $\delta\Phi$ satisfies the same equation as $\xi^r$. Reading off $\delta \Phi$ from the general solution of $F$ and labelling $\delta f=gf'$ and $\delta \tilde{f} = \tilde{g}\tilde{f}'$, the expression for $\xi^r$ reads:
\bea
\label{Liouvillediff1}
&&\xi^r=g' +\tilde{g}' +g\partial_+\Phi + \tilde{g}\partial_-\Phi,\cr
{\rm where}&&\partial_-g=\partial_+\tilde{g}=0.
\eea
We now use the covariant prescription prescribed in \cite{Barnich:2001jy} to compute the asymptotic conserved charges assiciated with the above diffeomorphisms.  
The infinitesimal change in the asymptotic charge under such diffeomorphisms is given by:
\be
\mathrlap{\slash}\delta Q =-\frac{l}{8\pi G} \int_{\partial {\mathcal M}}\! d\phi\, \left\{  2(\xi^+_{(0)}\delta\kappa + \xi^-_{(0)}\delta\tilde{\kappa})+ \frac{\delta F}{F^2}\xi^r(\partial_++\partial_-)F -\frac{\xi^r}{F}(\partial_++\partial_-)\delta F + \frac{\delta F}{F}(\partial_++\partial_-)\xi^r \right\}
\ee
The above charge is required to be integrable on the space of solutions. It can be shown to be so upto terms which vanish due to the integrand being a total derivative in  the angular co-ordinate $\phi$.
\be
\label{Liouvillecharge1}
\mathrlap{\slash}\delta Q = -\frac{l}{8\pi G}\int_{\partial {\mathcal M}}\!d \phi\, \left\{ 2(\xi^+_{(0)}\delta\kappa+\xi^-_{(0)}\delta\tilde{\kappa})+\delta\Phi(\partial_++\partial_-)\xi^r-\xi^r(\partial_++\partial_-)\delta\Phi \right\}.
\ee
The total integrated charge can be written again (after similarly throwing away total derivatives in $\phi$):
\be
Q = -\frac{l}{4\pi G}\int_{\partial {\mathcal M}}\!d\phi\,\left\{   \xi^+_{(0)}\kappa+\xi^-_{(0)}\tilde{\kappa}+ g(\partial^2_+\Phi-\tfrac{1}{2}(\partial_+\Phi)^2)+\tilde{g}(\partial^2_-\Phi-\tfrac{1}{2}(\partial_-\Phi)^2) \right\}. 
\ee
The factors multiplying $g$ and $\tilde{g}$ can be recognized as the stress-tensor modes of the Liouville theory. One can proceed to construct the classical Poisson brackets by demanding that the above charge gives rise to the fluctuations of the metric components $F$, $\kappa$ and $\tilde{\kappa}$ produced by the residual diffeomorphisms (\ref{residual_diff_Liouville}). The change in the parameters under such boundary condition preserving gauge transformations are: 
\bea
\label{Liouvillevariations}
\delta F&=& 2F\xi^r +\partial_+(F\xi^+_{(0)})+\partial_-(F\xi^-_{(0)}),\cr
\delta f&=& (g +\tfrac{1}{2}\xi^+_{(0)})f',\cr
\delta \tilde{f} &=& (\tilde{g} + \tfrac{1}{2}\xi^-_{(0)})\tilde{f}',\cr
\delta \kappa&=& \xi^+_{(0)}\kappa'+2 \kappa{\xi^+_{(0)}}'+ g''' +g'\hat{f} +\tfrac{1}{2}g\partial_+\hat{f},\cr
\delta \tilde{\kappa}&=&\xi^-_{(0)}\tilde{\kappa}'+2\tilde{\kappa}{\xi^-_{(0)}}'+\tilde{g}'''+\tilde{g}'\hat{\tilde{f}}+\tfrac{1}{2}\tilde{g}\partial_-\hat{\tilde{f}}.
\eea
Where $\hat{f}=2\partial^2_+\Phi-(\partial_+\Phi)^2$ and $\hat{\tilde{f}}=2\partial^2_-\Phi-(\partial_-\Phi)^2$. The variation in $F$ can be cast in terms of $\hat{f}$ and $\hat{\tilde{f}}$ as:
\bea
\tfrac{1}{2}\delta\hat{f}&=&(2g+\xi^+_{(0)})'''+(2g+\xi^+_{(0)})'\hat{f}+\tfrac{1}{2}(2g+\xi^+_{(0)})\partial_+\hat{f},\cr
\tfrac{1}{2}\delta\hat{\tilde{f}}&=&(2\tilde{g}+\xi^-_{(0)})'''+(2\tilde{g}+\xi^-_{(0)})'\hat{\tilde{f}}+\tfrac{1}{2}(2\tilde{g}+\xi^-_{(0)})\partial_-\hat{\tilde{f}}.
\eea
Therefore the space of classical solutions allowed by the proposed boundary condition(\ref{Liouville_bndy_cond}) are parametrized by the functions ($\hat{f},\hat{\tilde{f}},\kappa,\tilde{\kappa}$), where as the diffeomorphisms that would keep the metric under Lie derivative in this form are parametrized by ($g,\tilde{g},\xi^+_{(0)},\xi^-_{(0)}$). Redefining functions as:
\bea
\hat{f}\rightarrow \tfrac{1}{2}\hat{f},&&\hat{\tilde{f}}\rightarrow \tfrac{1}{2}\hat{\tilde{f}},\cr&&\cr
\kappa\rightarrow (\kappa-\tfrac{1}{2}\hat{f}),&&\tilde{\kappa}\rightarrow (\tilde{\kappa}-\tfrac{1}{2}\hat{\tilde{f}}),\cr&&\cr
g\rightarrow (g+\tfrac{1}{2}\xi^+_{(0)}),&&
\tilde{g}\rightarrow(\tilde{g}+\tfrac{1}{2}\xi^-_{(0)}),
\eea
the Poisson algebra reads:
\bea
-\tfrac{k}{2\pi}\left\{ \kappa({x^+}'),\kappa(x^+) \right\}&=&-[\kappa(x^+)+\kappa({x^+}')]\delta'({x^+}'-x^+)+\delta'''({x^+}'-x^+),\cr
-\tfrac{k}{2\pi} \left\{ \hat{f}({x^+}'),\hat{f}(x^+) \right\}&=&-[\hat{f}(x^+)+\hat{f}({x^+}')]\delta'({x^+}'-x^+)-\delta'''({x^+}'-x^+),\cr
\left\{ \hat{f}({x^+}',\kappa(x^+)) \right\}&=&0.
\eea  
Similarly for the left sector, which commutes with the right sector.
This shows that central charge associated with the Virasoros of the Liouville theory to be negative of the central charge of the Virasoros obtained from the Brown-Henneaux boundary conditions. The space of bulk geometries allowed by the Brown-Henneaux boundary conditions is contained in the space of solutions allowed by the above boundary condition.

\paragraph*{} If one begins with a generic 3d asymptotically locally $AdS_3$ metric in the Fefferman and Graham gauge then the residual diffeomorphisms are the ones which would generate the Diff$\times$Weyl\footnote{This is actually a semi-direct product where the commutator of Diff with Weyl is a Weyl.} for the boundary metric. Restricting the boundary metric to have the form as the one in \cite{Avery:2013dja} restricts the residual diffeomorphisms further to have an algebra chiral-Diff$\times$Witt to the leading order. Sub-leading order corrections to the residual diffeomorphisms further restrict it to an $sl(2,{\mathbb R})\times$Virasoro. Similarly, if on the other hand one imposed the above boundary conditions then the Diff$\times$Weyl reduces to two copies of left-right Virasoro with opposite central charges.   

The analyses here were done in the second order formalism of gravity using the Einstein-Hilbert action. For the sake of completeness next we look at the same problem from the first order formulation of $AdS_3$ gravity. 

\subsection{Liouville boundary conditions in CS formulation}
\label{appendixLiouvilleCS}
For the case of $AdS_3$ the gauge algebra of the CS theories is $sl(2,\mathbb{R})$. 
\be
S_{AdS_3}=S_{cs}[A]-S_{cs}[\tilde{A}]+S_{bndy},
\ee
where
\be S_{cs}[A]=\tfrac{k}{4\pi}\int tr(A\wedge dA+\tfrac{2}{3}A).\ee  
Here $k=\ell/4G$.
We first find the gauge fields which yield the metric proposed in (\ref{Liouville_bndy_cond}). As before, one mods out the radial $r$ dependence with a finite gauge transformation,
\bea
A&=&b^{-1}ab+b^{-1}db,\cr
\tilde{A}&=&b\tilde{a}b^{-1}+bdb^{-1},\cr
b&=&e^{\log\tfrac{r}{l}L_0}.
\eea
Since the equations of motion for $A$ and $\tilde{A}$ are flatness of their connections, one can equivalently work with $a$ and $\tilde{a}$ for the rest of the analysis.
Let us specialise the solution to be of the form:
\bea
a = (a^{(+)}_+ \, L_1 - a^{(-)}_+ \, L_{-1} + a^{(0)}_+ \, L_0) \, dx^+ + ( - a^{(-)}_- \, L_{-1} \, + a^{(0)}_- \, L_0) \, dx^-,
\eea
where $\{L_1,L_0,L_{-1}\}$ are the genrators of $sl(2,\mathbb{R})$\footnote{Here, $[L_m,L_n]=(m-n)L_{m+n}$.}.
Assuming that $a^{(+)}_+$ does not vanish, the flatness conditions imply:
\bea
a^{(0)}_- &=& \frac{1}{a^{(+)}_+} \partial_- a^{(+)}_+ , ~~
a^{(+)}_+ a^{(-)}_- = - \frac{1}{2} ( \partial_- a^{(0)}_+ - \partial_+a^{(0)}_-) \\
a^{(+)}_+ \, a^{(-)}_+ &=&  \kappa_0 (x^+) -\frac{1}{4} (a^{(0)}_+)^2 - \frac{1}{2} \partial_+ a^{(0)}_+ + \frac{1}{2} a^{(0)}_+ \, \partial_+ \ln a^{(+)}_+ + \frac{1}{2} \partial_+^2 \ln a^{(+)}_+- \frac{1}{4} (\partial_+ \ln a^{(+)}_+)^2 \nonumber
\eea
Similarly if we consider the 1-form
\bea
\tilde a = (\tilde a^{(+)}_+ \, L_1  + \tilde a^{(0)}_+ \, L_0) \, dx^+ + (- \tilde a^{(+)}_- \, L_1 + \tilde a^{(-)}_- \, L_{-1} \, + \tilde a^{(0)}_- \, L_0) \, dx^-
\eea
Then, assuming now that $\tilde a^{(-)}_-$ does not vanish, the flatness conditions read
\bea
\tilde a^{(-)}_-\tilde a^{(+)}_+ &=& - \frac{1}{2}( \partial_- \tilde a^{(0)}_+ - \partial_+\tilde a^{(0)}_-), ~~ \tilde a^{(0)}_+ = - \frac{1}{\tilde a^{(-)}_-} \, \partial_+\tilde a^{(-)}_- \\
\tilde a^{(-)}_- \, \tilde a^{(+)}_- &=& \tilde \kappa_0 (x^-) - \frac{1}{4} (\tilde a^{(0)}_-)^2 + \frac{1}{2} \partial_-\tilde a^{(0)}_- - \frac{1}{2} \tilde a^{(0)}_- \, \partial_- \ln \tilde a^{(-)}_- + \frac{1}{2} \partial_-^2 \ln \tilde a^{(-)}_- - \frac{1}{4} (\partial_- \ln \tilde a^{(-)}_-)^2 \nonumber 
\eea
The corresponding analysis in the second order formulation made use of the Fefferman-Graham (FG) gauge for the metric. One may impose this gauge on the above gauge fields by demanding that the metric corresponding to them be in the FG gauge. This is not strictly necessary but this has a benefit of reducing the number of solution space parameters by those ones which do not contribute to the asymptotic charge. Imposing the FG gauge on the metric translates to the following condition on the gauge field components:
\bea
a^{(0)}_+ = \tilde a^{(0)}_+, ~~ \tilde a^{(0)}_- = a^{(0)}_-
\eea
This gives the same metric as in (\ref{nlsoln2}) with the following identifications:
\bea
F = a^{(+)}_+ \tilde a^{(-)}_-, ~~ \kappa = a^{(+)}_+ a^{(-)}_+, ~~ \tilde \kappa = \tilde a^{(-)}_- \tilde a^{(+)}_-, ~~ \sigma = a^{(+)}_+ a^{(-)}_- = \tilde a^{(-)}_- \tilde a^{(+)}_+
\eea
\subsection{Asypmtotic symmetry analysis in the first order formalism}
Here we try and reproduce the results obtained in the second order formulation by starting out with the following gauge fields:
\bea
a&=&(a^{(+)}_+L_1-\partial_+(\log a^{(-)}_-)L_0-\tfrac{\kappa(x^+)}{a^{(+)}_+}L_{-1})dx^++(\partial_-(\log a^{(+)}_+)L_0-a^{(-)}_-L_{-1})dx^-,\cr
\tilde{a}&=&(\tilde{a}^{(+)}_+L_{-1}+\partial_-(\log \tilde{a}^{(+)}_+)L_0-\tfrac{\tilde{\kappa}(x^-)}{\tilde{a}^{(-)}_-}L_1)dx^-+(-\partial_+(\log \tilde{a}^{(-)}_-)L_0-\tilde{a}^{(+)}_+L_1)dx^+.
\eea
Above we have relabelled the parameters for the sake of computational convenience.
The above gauge fields reproduce the desired form of the metric with a FG constraint that $a^{(0)}=\tilde{a}^{(0)}$ along with the identifications
\be
F=-a^{(+)}_+\tilde{a}^{(-)}_-\,\, , \,\,\partial_+\partial_-\log(a^{(+)}_+a^{(-)}_-)=2a^{(+)}_+a^{(-)}_-\,\,,\,\,\partial_+\partial_-\log(\tilde{a}^{(+)}_+\tilde{a}^{(-)}_-)=2\tilde{a}^{(+)}_+\tilde{a}^{(-)}_-.
\ee
The residual gauge transformations are:
\bea
\Lambda&=&\Lambda^{(+)}a^{(+)}_+L_++\Lambda^{(0)}L_0+(-\tfrac{\kappa\Lambda^{(+)}}{a^{(+)}_+}+(y-\tilde{\Lambda}^{(-)})a^{(-)}_-)L_{-1},\cr
\tilde{\Lambda}&=&\tilde{\Lambda}^{(-)}\tilde{a}^{(-)}_-L_-+\tilde{\Lambda}^{(0)}L_0+(-\tfrac{\tilde{\kappa}\tilde{\Lambda}^{(-)}}{\tilde{a}^{(-)}_-}+(\tilde{y}-\Lambda^{(+)})\tilde{a}^{(+)}_+)L_+,\cr
{\rm where}\!\!&&\partial_-\Lambda^{(+)}=0=\partial_+\tilde{\Lambda}^{(-)}\,\,,\,\,\partial_-(a^{(+)}_+a^{(-)}_-\partial_+y)=0=\partial_+(\tilde{a}^{(+)}_+\tilde{a}^{(-)}_-\partial_-\tilde{y}).
\eea
The solutions to $y$ and $\tilde{y}$ can be given in terms of $\xi^r$ (\ref{Liouvillediff1}):
\be
y=-\frac{\partial_+\xi^r}{a^{(+)}_+a^{(-)}_-}\,\,,\,\,\tilde{y}=-\frac{\partial_-\xi^r}{\tilde{a}^{(+)}_+\tilde{a}^{(-)}_-}.
\ee
The FG constraint on the fluctuations yield the condition:
\be
\tilde{\Lambda}^{(0)}-\Lambda^{(0)}=\partial_-y+y\partial_-\log(a^{(+)}_+a^{(-)}_-)=\partial_+\tilde{y}+\tilde{y}\partial_+\log(\tilde{a}^{(+)}_+\tilde{a}^{(-)}_-).
\ee
Therefore the residual gauge transformation parameters are labelled by $\{g,\tilde{g},\Lambda^{(+)},\Lambda^{(-)}\}$. 
After imposing the FG gauge the on can write  
\be
F=\tfrac{1}{\chi}a^{(+)}_+a^{(-)}_-\,\,,\,\,\tilde{a}^{(+)}_+=-\tfrac{1}{\chi}a^{(+)}_+\,\,,\,\,\tilde{a}^{(-)}_-=-\chi a^{(-)}_-.
\ee
It turns out that the fluctuations of the gauge field components yield the same result for the metric components$\left\{ F,\kappa,\tilde{\kappa} \right\}$ as in the second order formalism with $\Lambda^{(+)}=\xi^+_{(0)}$ and $\Lambda^{(-)}=\xi^-_{(0)}$; explicitly given in (\ref{Liouvillevariations}).
The asymptotic charge for such configurations can then be written as:
\bea
\mathrlap{\slash}\delta Q&=&-\tfrac{k}{2\pi}\int\!d\phi\,(Tr[\Lambda.A_\phi]-Tr[\tilde{\Lambda}.\tilde{A}_\phi]),\cr
&=&\tfrac{\ell}{8\pi G}\int\,d\phi\,[\Lambda^{(+)}\delta\kappa+\tilde{\Lambda}^{(-)}\delta\tilde{\kappa}]\cr&&+\tfrac{(\tilde{\Lambda}^{(0)}-\Lambda^{(0)})}{2}\partial_-\delta\log a^{(+)}_+-\delta a^{(+)}_+a^{(-)}_-y +\tfrac{(\tilde{\Lambda}^{(0)}-\Lambda^{(0)})}{2}\partial_+\delta\log a^{(-)}_--\delta a^{(-)}_-a^{(+)}_+\tilde{y},\cr
&=&\tfrac{\ell}{16\pi G}\int\!d \phi\, \left\{ 2(\xi^+_{(0)}\delta\kappa+\xi^-_{(0)}\delta\tilde{\kappa})+\delta\Phi(\partial_++\partial_-)    \xi^r-\xi^r(\partial_++\partial_-)\delta\Phi \right\}.
\eea
The above expression for charge turns out to be the same as  (\ref{Liouvillecharge1}). Since the expressions for the fluctuations and the charges are the same in both the formalisms, we get the same asymptotic symmetry algebra as expected.


\bibliographystyle{utphys}

\begin{thebibliography}{10}

\bibitem{Brown:1986nw}
J.~D. Brown and M.~Henneaux, ``{Central Charges in the Canonical Realization of
  Asymptotic Symmetries: An Example from Three-Dimensional Gravity},''
\href{http://dx.doi.org/10.1007/BF01211590}{{\em Commun.Math.Phys.} {\bfseries
  104} (1986) 207--226}.

\bibitem{deBoer:1998kjm} 
  J.~de Boer,
  ``Six-dimensional supergravity on $S^3 \times AdS_3$ and 2-D conformal field theory,''
   \href{http://dx.doi.org/10.1016/S0550-3213(99)00160-1}{{\em   Nucl.\ Phys.\ B} {\bf 548}, 139 (1999)}
  [hep-th/9806104].
  

\bibitem{Ito:1998vd} 
  K.~Ito,
  ``Extended superconformal algebras on AdS(3),''
   \href{http://dx.doi.org/10.1016/S0370-2693(99)00070-2}{{\em  Phys.\ Lett.\ B} {\bf 449}, 48 (1999)}
  [hep-th/9811002].
  
\bibitem{Henneaux:1999ib} 
  M.~Henneaux, L.~Maoz and A.~Schwimmer,
  ``Asymptotic dynamics and asymptotic symmetries of three-dimensional extended AdS supergravity,''
  \href{http://dx.doi.org/10.1006/aphy.2000.5994}{{\em Annals Phys.}  {\bf 282}, 31 (2000)}
   [hep-th/9910013].
  

\bibitem{Banados:2002ey}
M.~Banados, O.~Chandia, and A.~Ritz, ``{Holography and the Polyakov action},''
  \href{http://dx.doi.org/10.1103/PhysRevD.65.126008}{{\em Phys.Rev.}
  {\bfseries D65} (2002) 126008},
\href{http://arxiv.org/abs/hep-th/0203021}{{\ttfamily arXiv:hep-th/0203021
  [hep-th]}}.


\bibitem{Compere:2008us}
G.~Compere and D.~Marolf, ``{Setting the boundary free in AdS/CFT},''
  \href{http://dx.doi.org/10.1088/0264-9381/25/19/195014}{{\em
  Class.Quant.Grav.} {\bfseries 25} (2008) 195014},
\href{http://arxiv.org/abs/0805.1902}{{\ttfamily arXiv:0805.1902 [hep-th]}}.

\bibitem{Troessaert:2013fma} 
  C.~Troessaert,
  ``Enhanced asymptotic symmetry algebra of $AdS$$_{3}$,''
  \href{http://dx.doi.org/10.1007/JHEP08(2013)044}{{\em JHEP} {\bf 1308}, 044 (2013)}  
  [arXiv:1303.3296 [hep-th]].

\bibitem{Avery:2013dja} 
  S.~G.~Avery, R.~R.~Poojary and N.~V.~Suryanarayana,
  ``An sl(2,$\mathbb{R}$) current algebra from $AdS_3$ gravity,''
  \href{http://dx.doi.org/10.1007/JHEP01(2014)144}{JHEP {\bf 1401}, 144 (2014)},
  \href{http://arxiv.org/abs/1304.4252}{\ttfamily arXiv:1304.4252 [hep-th]}.

\bibitem{rrp2015thesis}
Rohan Raghava Poojary, 
``Aspects of Holographic Induced Gravities",
\href{http://www.hbni.ac.in/students/dsp_ths.html?nm=phys/PHYS10200904001.pdf}{HBNI, PhD Thesis, July 2015} .


\bibitem{Apolo:2014tua} 
  L.~Apolo and M.~Porrati,
  ``Free boundary conditions and the AdS$_3$/CFT$_2$ correspondence,''
    \href{http://dx.doi.org/doi:10.1007/JHEP03(2014)116}{{\em JHEP} {\bf 1403}, 116 (2014)}
  [arXiv:1401.1197 [hep-th]].

 \bibitem{Compere:2013bya}
G.~Comp\`{e}re, W.~Song, and A.~Strominger, ``{New Boundary Conditions for
  AdS3},''
\href{http://arxiv.org/abs/1303.2662}{{\ttfamily arXiv:1303.2662 [hep-th]}}.


\bibitem{Grumiller:2016pqb} 
  D.~Grumiller and M.~Riegler,
  ``Most general AdS$_{3}$ boundary conditions,''
     \href{http://dx.doi.org/doi:10.1007/JHEP10(2016)023}{{\em JHEP} {\bf 1610}, 023 (2016)}
  [arXiv:1608.01308 [hep-th]].
   
\bibitem{Perez:2016vqo} 
  A.~Pérez, D.~Tempo and R.~Troncoso,
  ``Boundary conditions for General Relativity on AdS$_{3}$ and the KdV hierarchy,''
   \href{http://dx.doi.org/doi:10.1007/JHEP06(2016)103}{{\em JHEP} {\bf 1606}, 103 (2016)}
  [arXiv:1605.04490 [hep-th]].
  
\bibitem{Krishnan:2016dgy} 
  C.~Krishnan, A.~Raju and P.~N.~B.~Subramanian,
  ``Dynamical boundary for antiÐde Sitter space,''
   \href{http://dx.doi.org/doi:10.1103/PhysRevD.94.126011}{ {\em Phys.\ Rev.}\ D {\bf 94}, no. 12, 126011 (2016)}
  [arXiv:1609.06300 [hep-th]].

\bibitem{Polyakov:1981rd}
A.~M. Polyakov, ``{Quantum Geometry of Bosonic Strings},''
\href{http://dx.doi.org/10.1016/0370-2693(81)90743-7}{{\em Phys.Lett.}
  {\bfseries B103} (1981) 207--210}.

\bibitem{Polyakov:1987zb}
A.~M. Polyakov, ``{Quantum Gravity in Two-Dimensions},''
\href{http://dx.doi.org/10.1142/S0217732387001130}{{\em Mod.Phys.Lett.}
  {\bfseries A2} (1987) 893}.

\bibitem{Grisaru:1987pf} 
  M.~T.~Grisaru and R.~M.~Xu,
  ``Quantum Supergravities in Two-dimensions,''
  \href{http://dx.doi.org/10.1016/0370-2693(88)90983-5}{{\em Phys.\ Lett.\ B} {\bf 205}, 486 (1988)}.

\bibitem{HariDass:1988dp} 
  N.~D.~Hari Dass and R.~Sumitra,
  ``Symmetry Reorganization in Exactly Solvable Two-dimensional Quantized Supergravity,''
   \href{http://dx.doi.org/10.1142/S0217751X8900090X}{{\em  Int.\ J.\ Mod.\ Phys.\ A} {\bf 4}, 2245 (1989)}.
 
\bibitem{Bondi:1962px} 
  H.~Bondi, M.~G.~J.~van der Burg and A.~W.~K.~Metzner,
  ``Gravitational waves in general relativity. 7. Waves from axisymmetric isolated systems,''
    \href{http://dx.doi.org/doi:10.1098/rspa.1962.0161}{{\em  Proc.\ Roy.\ Soc.\ Lond.}\ A {\bf 269}, 21 (1962)}.

\bibitem{Sachs:1962zza} 
  R.~Sachs,
  ``Asymptotic symmetries in gravitational theory,''
    \href{http://dx.doi.org/doi:10.1103/PhysRev.128.2851}{{\em Phys.\ Rev.}\  {\bf 128}, 2851 (1962)}.

\bibitem{Banados:2004nr} 
  M.~Banados and R.~Caro,
  ``Holographic ward identities: Examples from 2+1 gravity,''
    \href{http://dx.doi.org/10.1088/1126-6708/2004/12/036}{{\em JHEP} {\bf 0412}, 036 (2004)}
  [hep-th/0411060].

\bibitem{Barnich:2012aw} 
  G.~Barnich, A.~Gomberoff and H.~A.~Gonzalez,
  ``The Flat limit of three dimensional asymptotically anti-de Sitter spacetimes,''
   \href{http://dx.doi.org/10.1103/PhysRevD.86.024020}{{\em  Phys.\ Rev.\ D} {\bf 86}, 024020 (2012)}
  [arXiv:1204.3288 [gr-qc]].
  
\bibitem{Barnich:2014cwa} 
  G.~Barnich, L.~Donnay, J.~Matulich and R.~Troncoso,
  ``Asymptotic symmetries and dynamics of three-dimensional flat supergravity,''
    \href{http://dx.doi.org/10.1007/JHEP08(2014)071}{{\em JHEP} {\bf 1408}, 071 (2014)}
[arXiv:1407.4275 [hep-th]].
            

\bibitem{Banerjee:2016nio} 
  N.~Banerjee, D.~P.~Jatkar, I.~Lodato, S.~Mukhi and T.~Neogi,
  ``Extended Supersymmetric BMS$_3$ algebras and Their Free Field Realisations,''
    \href{http://dx.doi.org/10.1007/JHEP11(2016)059}{{\em JHEP} {\bf 1611}, 059 (2016)}
  [arXiv:1609.09210 [hep-th]].

\bibitem{Banerjee:2017gzj} 
  N.~Banerjee, I.~Lodato and T.~Neogi,
  ``N=4 Supersymmetric BMS3 algebras from asymptotic symmetry analysis,''
     \href{http://dx.doi.org/10.1103/PhysRevD.96.066029}{{\em Phys.\ Rev.\ D} {\bf 96}, no. 6, 066029 (2017)}, 
  [arXiv:1706.02922 [hep-th]].

\bibitem{Fuentealba:2017fck}
  O.~Fuentealba, J.~Matulich and R.~Troncoso,
  JHEP {\bf 1709} (2017) 030
  doi:10.1007/JHEP09(2017)030
  [arXiv:1706.07542 [hep-th]].

  
\bibitem{Poojary:2014ifa} 
  R.~R.~Poojary and N.~V.~Suryanarayana,
  ``Holographic chiral induced W-gravities,''
  \href{http://dx.doi.org/10.1007/JHEP10(2015)168}{JHEP {\bf 1510}, 168 (2015)},
  \href{http://arxiv.org/abs/1412.2510}{\ttfamily arXiv:1412.2510 [hep-th]}.
  
\bibitem{Achucarro:1987vz} 
  A.~Achucarro and P.~K.~Townsend,
  ``A Chern-Simons Action for Three-Dimensional anti-De Sitter Supergravity Theories,''
   \href{http://dx.doi.org/doi:10.1016/0370-2693(86)90140-1}{{\em  Phys.\ Lett.\ B} {\bf 180}, 89 (1986)}.
  
\bibitem{Witten:1988hc} 
  E.~Witten,
  ``(2+1)-Dimensional Gravity as an Exactly Soluble System,''
  Nucl.\ Phys.\ B {\bf 311}, 46 (1988).

\bibitem{Bautier:1999ds} 
  K.~Bautier,
  ``AdS(3) asymptotic (super)symmetries,''
  hep-th/9909097.

\bibitem{Banados:1998pi} 
  M.~Banados, K.~Bautier, O.~Coussaert, M.~Henneaux and M.~Ortiz,
  ``Anti-de Sitter / CFT correspondence in three-dimensional supergravity,''
    \href{http://dx.doi.org/10.1103/PhysRevD.58.085020}{{\em Phys.\ Rev.\ D} {\bf 58}, 085020 (1998)
}
  [hep-th/9805165].

\bibitem{Barnich:2001jy}
G.~Barnich and F.~Brandt, ``{Covariant theory of asymptotic symmetries,
  conservation laws and central charges},''
  \href{http://dx.doi.org/10.1016/S0550-3213(02)00251-1}{{\em Nucl.Phys.}
  {\bfseries B633} (2002) 3--82},
\href{http://arxiv.org/abs/hep-th/0111246}{{\ttfamily arXiv:hep-th/0111246
  [hep-th]}}.

\bibitem{Barnich:2007bf}
G.~Barnich and G.~Comp\`{e}re, ``{Surface charge algebra in gauge theories and
  thermodynamic integrability},''
  \href{http://dx.doi.org/10.1063/1.2889721}{{\em J.Math.Phys.} {\bfseries 49}
  (2008) 042901},
\href{http://arxiv.org/abs/0708.2378}{{\ttfamily arXiv:0708.2378 [gr-qc]}}.

\bibitem{Nahm:1977tg} 
  W.~Nahm,
  ``Supersymmetries and their Representations,''
    \href{http://dx.doi.org/10.1016/0550-3213(78)90218-3}{{\em Nucl.\ Phys.\ B} {\bf 135}, 149 (1978)}.


\bibitem{Bagchi:2009my} 
  A.~Bagchi and R.~Gopakumar,
  ``Galilean Conformal Algebras and AdS/CFT,''
  \href{http://dx.doi.org/doi:10.1088/1126-6708/2009/07/037}{ {\em JHEP} {\bf 0907}, 037 (2009)}
  [arXiv:0902.1385 [hep-th]].
  
\bibitem{Knizhnik:1986wc} 
  V.~G.~Knizhnik,
  ``Superconformal Algebras in Two-dimensions,''
  \href{http://dx.doi.org/10.1007/BF01028940}{{\em Theor.\ Math.\ Phys.}\  {\bf 66}, 68 (1986)}
  [Teor.\ Mat.\ Fiz.\  {\bf 66}, 102 (1986)].

\bibitem{Bershadsky:1986ms} 
  M.~A.~Bershadsky,
  ``Superconformal Algebras in Two-dimensions With Arbitrary $N$,''
  \href{http://dx.doi.org/10.1016/0370-2693(86)91100-7}{{\em   Phys.\ Lett.\ B} {\bf 174}, 285 (1986)}.
  
\bibitem{Schoutens:1988tg} 
  K.~Schoutens,
  ``Representation Theory for a Class of SO($N$) Extended Superconformal Operator Algebras,''
    \href{http://dx.doi.org/10.1016/0550-3213(89)90163-6}{{\em Nucl.\ Phys.\ B} {\bf 314}, 519 (1989)}.
 
\bibitem{Defever:1991tf} 
  F.~Defever, W.~Troost and Z.~Hasiewicz,
  ``Superconformal algebras with quadratic nonlinearity,''
   \href{http://dx.doi.org/10.1016/0370-2693(91)90552-2}{{\em  Phys.\ Lett.\ B} {\bf 273}, 51 (1991)}.


\bibitem{Fradkin:1991gj} 
  E.~S.~Fradkin and V.~Y.~Linetsky,
  ``An Exceptional N=8 superconformal algebra in two-dimensions associated with F(4),''
  \href{http://dx.doi.org/10.1016/0370-2693(92)91600-E}{{\em   Phys.\ Lett.\ B} {\bf 275}, 345 (1992)}.
  
\bibitem{Fradkin:1992km} 
  E.~S.~Fradkin and V.~Y.~Linetsky,
  ``Classification of superconformal and quasisuperconformal algebras in two-dimensions,''
   \href{http://dx.doi.org/10.1016/0370-2693(92)90120-S}{{\em   Phys.\ Lett.\ B} {\bf 291}, 71 (1992)}.
  
\bibitem{Fradkin:1992bz} 
  E.~S.~Fradkin and V.~Y.~Linetsky,
  ``Results of the classification of superconformal algebras in two-dimensions,''
   \href{http://dx.doi.org/10.1016/0370-2693(92)90651-J}{{\em  Phys.\ Lett.\ B} {\bf 282}, 352 (1992)}
  [hep-th/9203045].
  
\bibitem{Fradkin:1992cb} 
  E.~S.~Fradkin and V.~Y.~Linetsky,
  ``Classification of superconformal algebras with quadratic nonlinearity,''
  \href{http://arxiv.org/abs/hep-th/9207035}{{\ttfamily arXiv:hep-th/9207035}}.
  
\bibitem{Bowcock:1992bm} 
  P.~Bowcock,
  ``Exceptional superconformal algebras,''
   \href{http://dx.doi.org/10.1016/0550-3213(92)90654-T}{{\em   Nucl.\ Phys.\ B} {\bf 381}, 415 (1992)}
  [hep-th/9202061].

             
\bibitem{Bina:1997bm} 
  B.~Bina and M.~Gunaydin,
  ``Real forms of nonlinear superconformal and quasisuperconformal algebras and their unified realization,''
     \href{http://dx.doi.org/10.1016/S0550-3213(97)00406-9}{{\em Nucl.\ Phys.\ B} {\bf 502}, 713 (1997)}
  [hep-th/9703188].
  
\bibitem{Ishimoto:1998hd} 
  Y.~Ishimoto,
  ``Classical Hamiltonian reduction on $D(2|1, \alpha)$ Chern-Simons gauge theory and large N=4 superconformal symmetry,''
    \href{http://dx.doi.org/10.1016/S0370-2693(99)00624-3}{{\em Phys.\ Lett.\ B} {\bf 458}, 491 (1999)}
  [hep-th/9808094].

  
\bibitem{Kraus:2007vu} 
  P.~Kraus, F.~Larsen and A.~Shah,
  ``Fundamental Strings, Holography, and Nonlinear Superconformal Algebras,''
    \href{http://dx.doi.org/10.1088/1126-6708/2007/11/028}{{\em JHEP} {\bf 0711}, 028 (2007)}
  [arXiv:0708.1001 [hep-th]].
  
\bibitem{Grumiller:2017sjh} 
  D.~Grumiller, W.~Merbis and M.~Riegler,
  ``Most general flat space boundary conditions in three-dimensional Einstein gravity,''
     \href{http://dx.doi.org/doi:10.1088/1361-6382/aa8004}{{\em Class.\ Quant.\ Grav.}  {\bf 34}, no. 18, 184001 (2017)}
  [arXiv:1704.07419 [hep-th]].
                      
\bibitem{Fuentealba:2017omf} 
  O.~Fuentealba, J.~Matulich, A.~Pérez, M.~Pino, P.~Rodríguez, D.~Tempo and R.~Troncoso,
  ``Integrable systems with BMS$_{3}$ Poisson structure and the dynamics of locally flat spacetimes,''
  \href{http://arxiv.org/abs/1711.02646}{arXiv:1711.02646 [hep-th]}.
  
  
\bibitem{Skenderis:1999nb}
K.~Skenderis and S.~N. Solodukhin, ``{Quantum effective action from the AdS /
  CFT correspondence},''
  \href{http://dx.doi.org/10.1016/S0370-2693(99)01467-7}{{\em Phys.Lett.}
  {\bfseries B472} (2000) 316--322},
\href{http://arxiv.org/abs/hep-th/9910023}{{\ttfamily arXiv:hep-th/9910023
  [hep-th]}}.




















 
  
  
  
  
  
  
  
  









  






  
    
  
 
   


\end{thebibliography}
\providecommand{\href}[2]{#2}\begingroup\raggedright\endgroup
\end{document}